\documentclass[3p,10pt,twoside,final,authoryear]{elsarticle}

\usepackage[latin1]{inputenc}

\usepackage{amsmath,amsthm,amssymb,bm}
\usepackage{graphicx}
\usepackage{mathptmx}
\usepackage{tikz}
\usetikzlibrary{calc,patterns,decorations.pathmorphing,decorations.markings}
\usepackage[colorlinks,linkcolor=blue,citecolor=blue,urlcolor=blue]{hyperref}
\usepackage{natbib}

\makeatletter
\renewcommand\@makefntext[1]{\leftskip=0.0em\hskip-0.5em\@makefnmark{#1}}
\def\ps@pprintTitle{
   \let\@oddhead\@empty
   \let\@evenhead\@empty
   \def\@oddfoot{\footnotesize
       \it Accepted for publication in Icarus (Elsevier) \hfill December 18, 2019}
   \let\@evenfoot\@oddfoot}
\makeatother

\usepackage{anyfontsize}

\begin{document}

\begin{frontmatter}

\title{Andrade rheology in time-domain. Application to Enceladus' dissipation of energy due to forced libration}

\author[obs,ime]{Yeva Gevorgyan} \ead{yeva@ime.usp.br}
\author[obs]{Gwena\"{e}l Bou\'{e}} \ead{gwenael.boue@obspm.fr}
\author[ime]{Clodoaldo Ragazzo} \ead{ragazzo@ime.usp.br}
\author[unifei]{Lucas S. Ruiz} \ead{lucasruiz@unifei.edu.br}
\author[UA]{Alexandre C.~M. Correia}\ead{acor@uc.pt}

\address[obs]{ASD/IMCCE,  CNRS-UMR8028,  Observatoire  de  Paris,  PSL  University,  Sorbonne  Universit{\'e},  77  Avenue  Denfert-Rochereau, 75014 Paris, France}
\address[ime]{Instituto de Matem\'{a}tica e Estat\'{i}stica, Universidade de S\~{a}o Paulo, 05508-090 S\~{a}o Paulo, SP, Brazil}
\address[unifei]{Instituto de Matem\'{a}tica e Computa\c{c}\~{a}o, Universidade Federal de Itajub\'{a}, 37500-903 Itajub\'{a}, MG, Brazil}
\address[UA]{CFisUC, Department of Physics, University of Coimbra, 3004-516 Coimbra, Portugal}


\begin{abstract}
  The main purpose of this work is to present a time-domain implementation of the Andrade rheology,
  instead of the traditional expansion in terms of a Fourier series of the tidal potential. This
  approach can be used in any fully three dimensional numerical simulation of the dynamics of a system
  of many deformable bodies. In particular, it allows large eccentricities, large mutual inclinations,
  and it is not limited to quasi-periodic perturbations. It can take into account an extended class of
  perturbations, such as chaotic motions, transient events, and resonant librations.

  The results are presented by means of a concrete application:  the analysis of the libration
  of Enceladus. This is done  by means of both analytic formulas in the frequency domain
and direct numerical simulations.
  We do not   a priori  assume that Enceladus has a triaxial shape,
  the eventual triaxiality  is a consequence of the satellite motion and its  rheology. As a result
  we obtain an analytic formula for the amplitude of libration  that incorporates a new correction due to
  the rheology.

  Our results provide an estimation of the amplitude of libration of the core of Enceladus as 0.6\%
  of that of the shell. They also reproduce the observed 10\,GW of tidal heat generated by
  Enceladus with a value of $0.17\times 10^{14}$Pa$\cdot$s for the global effective viscosity under
  both Maxwell and Andrade rheology.

\end{abstract}



\end{frontmatter}


\section{Introduction}

\citet{and1910} observed that metallic wires subject to a constant stress tend to stretch as a function of time $t$, according to a law in $t^{1/3}$. He attributed this phenomenon to a rearrangement and/or a rotation of small parts of a crystalline structure switching between successive equilibrium states. It has
been shown to accurately model the transient deformation of many metals, but also of polycrystalline
ice \citep[and references therein]{Glen1955,McCarthy2013}. A theoretical derivation of
the 1/3 power has been proposed by \citet{Mott1953}. At high pressure and temperature conditions,
rock and rock-forming minerals also behave like metals \citep{Griggs1960}. Laboratory experiments on
rock periodically stressed at frequencies spanning $10^{-4}$ to $10^1$\,rad/s reproduce an Andrade
like creep rheology with a power $\alpha$ varying as a function of the composition between 0.33 and
0.59 \citep[e.g.][]{Goetze1971, Goetze1972}.

The application of Andrade's rheology in problems of convection in the Earth mantle has been
discussed by \citet{Stacey1963}. However this law can only be valid over a finite range of
frequencies otherwise it would prevent thermal convection and continental drift
\citep{Jeffreys1972}. Geophysical data stemming from body waves, surface waves, free oscillations, and
Chandler wobble, covering a large frequency domain also suggest that the mantle dissipates energy
according to $Q^{-1} \propto \omega^{-\alpha}$, where $Q$ is the quality factor, $\omega$ the
excitation frequency and $\alpha$ a coefficient between 1/5 and 1/3, in agreement with Andrade's
creep model \citep{Anderson1979}. This behaviour is observed in the lower mantle for $\omega$ in the
range $10^{-8}$ -- $10^{-1}$\,rad/s and in the upper mantle for $\omega$ in $10^{-3}$ --
$10^{2}$\,rad/s. For lower and greater frequencies, the expected dissipation rates are $Q^{-1}
\propto \omega$ and $Q^{-1} \propto \omega^{-1}$, respectively (ibid.). \citet{Anderson1979}
attribute the plateau feature at intermediary frequencies as the consequence of a continuum spectrum
of retardation times, which turns out to be a physical motivation for the construction of
the approximate Andrade model presented in Section~\ref{aprox}. The width of the frequency band
where $Q^{-1}$ evolves as $\omega^{-\alpha}$ is related to the range of dislocation density and grain
size involved in grain boundary processes and defect motions \citep{Karato1990}.
 More recent geophysical data suggest
for the Earth an exponent $\alpha$ of the order of 0.15 \citep{Petit2010} and 
an upper frequency of the plateau $Q^{-1} \propto \omega^{-\alpha}$ in the Earth mantle around
$10^{-3}$\,rad/s \citep{Lau2019}.

The Andrade creep model has been introduced in the astronomical community by \citet{Efroimsky2007}
and \citet{efr2012, Efroimsky2012}. In comparison with more standard rheologies, it can
significantly affect the probability of capture in spin-orbit resonance and/or amplify the amount of
dissipated energy \citep{Makarov2013, Leconte2015, Makarov2015, Makarov2016,
Walterova2017, Renaud2018}. The Andrade rheology has been applied in studies of specific celestial
bodies such as the Moon \citep{Nimmo2012, Williams2014, Williams2015}, Mercury \citep{Padovan2014,
Noyelles2014, Knibbe2017}, Enceladus \citep{Rambaux2010, SHK2013, Behounkova2013, Behounkova2015,
Soucek2019}, Iapetus \citep{CastilloRogez2011}, Io \citep{Bierson2016}, binary asteroids
\citep{Efroimsky2015}, GJ581\,d \citep{Makarov2012}, and Proxima Century\,b \citep{Ribas2016}.

The main objective of this paper is to present finite dimensional sets of ordinary differential equations that approximate the Andrade rheology with an arbitrary degree of accuracy.
Many of the visco-elastic models used in geophysics, Kelvin-Voigt, Maxwell, Burgers, etc, can easily
be written in the time domain allowing for applications in cases of large eccentricities or when
perturbations are not quasi-periodic \citep{FerrazMello2013, refId0, FerrazMello2015,
FerrazMello2015b, bcl2016, Folonier2018}.
In the frequency domain the only difference of the Andrade model from the others is the appearance
of a fractional power of the frequency in the complex compliance function. While in the frequency
domain this difference is innocuous, in time domain it becomes quite challenging. Indeed, while
integral power laws of the frequency are related to ordinary derivatives, the fractional power laws
are related to fractional derivatives, which are integro-differential operators. For this reason, as
far as we know, equations of motion that describe simultaneously the rotation, the position, and
tide deformation under Andrade rheology were never considered in the time domain. Nevertheless, as
suggested by \citet{Anderson1979}, Andrade's fractional power law can be interpreted as a continuum
of Voigt elements in series, which can itself be approximated by a discrete series of Voigt elements
\citep[e.g.,][]{Birk2010, BenJazia2013}. The resulting model, which is known as the extended Burgers
model, allows to describe any rheology in the time domain.
 
To validate the method we analyse the energy dissipation of Enceladus due to the libration which
naturally appears in the simulations. In 
this framework this effect can indeed be independently estimated analytically.
This goal allowed us to integrate, for the first time, the equations of motion of a celestial body, Enceladus,  using the Andrade rheology to describe tidal deformations.

Enceladus is one of the medium sized moons of Saturn with a radius of 252.1 km, mostly covered by ice. Currently it is synchronised with its host planet. It is orbiting the planet in a slightly eccentric orbit ($e=0.0045$) due to it being in $2:1$ mean motion resonance with another satellite of Saturn, Dione. During the flyby of July 14th, 2005, the Cassini Composite Infrared Spectrometer (CIRS) found plume activity around four large fractures near the south pole of Enceladus (see \cite{Porco1393}). Currently, it is believed that the plum activity on Enceladus is due to tidal heating. No appreciable radiogenic heating is believed to be involved in this process. The latest estimate for the energy flux from the fractures is around 10 GW \citep{KaN2017}. According to existing scenarios of satellite formation \cite{can2006, CCC2011, SaC2017, AsR2013} radioactive decay no longer has contribution to the generation of the heat on Enceladus (for details see \cite{efr2018}).

 The Andrade rheology has some parameters that must be fit from observations.
 In  \cite{efr2018}, using the Fourier expansion to compute the tidal energy dissipation,
 with additional terms accounting for the forced longitudinal libration, the value found
 for the mean viscosity was $\eta \approx 0.24\times 10^{14}$ Pa$\cdot$s, which is expected for an ice shell.
 A range $\eta \approx (0.6-1.9)\times 10^{14}$ Pa$\cdot$s was estimated in \cite{Folonier2018}.
 Here we use $\eta = 0.17\times 10^{14}$\,Pa$\cdot$s both in the direct integrations of the unaverage
 equations of motion and in the analytical formulae.

The paper is organised as follows. 

In Section \ref{EDS}  we first construct non-dissipative equations of motion for the
Enceladus-Saturn system.  Motion on a fixed orbit is introduced to include the effect of the
Enceladus-Dione resonance.
In Section \ref{sec_generic_rheology} we present dissipation in the standard way in terms of the
Love number $k_2$.
Then in Section \ref{sec_rheology} the dissipation is introduced into the equations of motion by
means of the Association Principle given  in \cite{RaR2017}.  This is the key point in the
implementation of the Andrade rheology in time-domain.  Both Maxwell and Andrade models  are used to
describe the rheology of  Enceladus. In this way we can evaluate the difference between distinct
rheological models on the dissipation of energy. 
In Section \ref{aprox} we build up  on previous results about finite-dimensional approximations to
fractional derivative operators and present a finite dimensional approximation of the Andrade
rheology by means of an extended Burgers model.
Section \ref{sec_libration} is dedicated to forced librations. 
We present a new analytical
expression for the libration magnitude and another for the average energy dissipation rate in order
to validate the simulations. 
In Section \ref{sec_integration} we discuss the choice of parameters and initial conditions used in
the integrations and present our results. The dissipation rates obtained for different rheologies
are compared.
 In Section \ref{difference} we analyse the
difference between the angle of libration obtained from the model and that observed.  This
difference between the two angles and an interior model for Enceladus is used to estimate the
libration angle of Enceladus core.

In the Conclusion \ref{conclusions} we summarise the main results in the paper.


\section{Dynamical model}

\label{EDS}

Consider Enceladus as a deformable body orbiting a point-like Saturn. Let $\kappa$ be an inertial reference frame at the centre of mass of the system, which is at rest. Let $q^{\prime 1}\in\kappa$ and $q^{\prime 2}\in\kappa$ be the positions, and $m_1$ and $m_2$ the masses of Enceladus and Saturn, respectively. ${\rm I}_\circ$ is one third of the trace of the inertia tensor of Enceladus. We suppose that the deformations of the satellite are incompressible which implies that ${\rm I}_\circ$ is constant.

In order to describe the rotation of Enceladus, 
let $\mathbf K$ be a body frame of Enceladus with origin at its centre of mass. $\mathbf K$ is defined as being a Tisserand frame, namely
with respect to $\mathbf K$ the angular momentum of Enceladus is null \citep{MunkMac}. Let 
$\mathbf Y: \mathbf K\to\kappa$ be the orientation matrix associated with $\mathbf K$ and
$\boldsymbol {\Omega}={\mathbf Y}^{-1}\dot{\mathbf Y}:\mathbf K\to\mathbf K$ be the angular velocity operator.
Notice that  $\mathbf Y$,
and therefore $\mathbf K$, will be determined by the integration of the equations of motion, they are not known a priori.

The deformations of Enceladus imply that 
the moment of inertia operator $\mathbf I:\mathbf K\to\mathbf K$ is not constant in time. It can be written as
$\mathbf I= {\rm I}_\circ(\mathbb{I}-\mathbf{B})$, where $\mathbb I$ is the identity matrix and $\mathbf{B}$  
is a  symmetric traceless matrix that encodes all the deformation of Enceladus.

Let $q={\mathbf Y}^{-1}(q^{\prime 2}-q^{\prime 1})$ be the relative position of the planet in the reference frame $\mathbf K$.
The equations for the orbital motion and the spin of the extended body,
in the body frame, are (for details see \cite{RaR2017}), 
\begin{equation}\label{eqmot}
  \begin{split}
    \frac{\dot{\mathbf{L}}}{{\rm I}_{\circ}}&=\dot{\boldsymbol{\Omega}}+\dot{\mathbf{B}}\boldsymbol{\Omega}+\boldsymbol{\Omega}\dot{\mathbf{B}}
    +\mathbf{B}\dot{\boldsymbol{\Omega}}+\dot{\boldsymbol{\Omega}} \mathbf{B}\\
    &=[\mathbf{B},\boldsymbol{\Omega}^2]+\frac{3G m_{2}}{|q|^{5}}\left[q\otimes q,\mathbf{B}\right]\\
\dot{q}&= -\boldsymbol{\Omega}q+v \\
\dot{v}&= -\boldsymbol{\Omega}v
+ G(m_1+m_2)\bigg\{
 -\frac{1}{|q|^{3}}q
\\ &
-\frac{15}{2}\frac{{\rm I}_{\circ}}{m_{1}}\frac{1}{|q|^{7}}(q\cdot\mathbf{B}q)q+3\frac{{\rm I}_{\circ}}{m_{1}}\frac{1}{|q|^{5}}
\mathbf{B}q\bigg\},
\end{split}
\end{equation}
where $G=6.6743\times 10^{-11}\mathrm{m}^3\cdot\mathrm{kg}^{-1}\cdot\mathrm{s}^{-2}$ is the gravitational constant,
$q$ and $v$ are respectively the position and the velocity vectors,
$[\mathbf{B},\boldsymbol{\Omega}]=\mathbf{B}\boldsymbol {\Omega}-\boldsymbol{\Omega}\mathbf{B}$ denotes
the commutator of matrices, and $q\otimes q$ denotes the matrix with entries $(q\otimes q)_{ij}=q_{i}q_{j}$.
The total angular momentum is given by
\begin{equation}
\mathbf{L}={\rm I}_{\circ}(\boldsymbol{\Omega}+\mathbf{B}\boldsymbol{\Omega}+\boldsymbol{\Omega}\mathbf{B}).
\end{equation}

To keep Enceladus on an eccentric orbit and to attain the forced oscillations regime we would have to add Dione to the model. Here we alternatively fix the eccentricity and the semi-major axis at the present values and remove tidal evolution from the orbits.

It is more convenient to integrate the equations in the inertial reference frame
\begin{equation}\label{inertfr}
q'=\mathbf{Y}q, \qquad v'=\mathbf{Y}v, \qquad \boldsymbol{\omega}=\mathbf{Y}\boldsymbol{\Omega}\mathbf{Y}^{T}, \qquad \mathbf{b}=\mathbf{Y}\mathbf{B}\mathbf{Y}^{T},
\end{equation}
where ${}^{T}\!$ denotes the transpose operator. Notice that for any matrix $\mathbf{x}=\mathbf{Y}\mathbf{X}\mathbf{Y}^T$, its time derivative satisfies
\begin{equation}\label{timeder}
\dot{\mathbf{X}}=\mathbf{Y}^T\dot{\mathbf{x}}\mathbf{Y}+\dot{\mathbf{Y}}^T\mathbf{x}\mathbf{Y}+\mathbf{Y}^T\mathbf{x}\dot{\mathbf{Y}}=\mathbf{Y}^T\Big(\dot{\mathbf{x}}+\mathbf{Y}\dot{\mathbf{Y}}^T\mathbf{x}+\mathbf{x}\dot{\mathbf{Y}}\mathbf{Y}^T\Big)\mathbf{Y}=\mathbf{Y}^T(\mathbf{x}-[\boldsymbol{\omega},\mathbf{x}])\mathbf{Y}.
\end{equation}
In the inertial frame the  equations of motion become
\begin{equation}\label{moteqn1}
\begin{split}
\dot{\boldsymbol{\omega}}+\dot{\mathbf{b}}\boldsymbol{\omega}+\boldsymbol{\omega}\dot{\mathbf{b}}+\mathbf{b}\dot{\boldsymbol{\omega}}+\dot{\boldsymbol{\omega}}\mathbf{b}&= \frac{3G m_{2}}{|q'|^{5}}\left[q'\otimes q',\mathbf{b}\right]\\
\dot{q}^\prime&= v'\\ 
\dot{v}^\prime&= -G(m_1+m_2)\frac{1}{|q'|^3}q'\,.
\end{split}
\end{equation}
Equations for $\mathbf{b}$, which are given in the next two sections, are still necessary to close the system.


\section{Tidal equations in the frequency domain}\label{sec_generic_rheology}

Let $V'(z, t)$ be the additional quadrupolar potential  at time $t$ associated with the deformation of Enceladus
at a point $z\in \mathbf{K}$ of its surface, i.e. such that $|z| = R$, where $R$ is the Enceladus' mean radius, and
induced by the perturbing quadrupolar potential $W(z, t)$. By definition of the Love number $k_2$,
and according to the equivalence principle \citep[Section 6.3]{EFR2018328},
\begin{equation}\label{LoveVW}
\hat V'(z, \omega) = k_2(\omega) \hat W(z, \omega)\ ,
\end{equation}
where $\ \hat{}\ $ denotes the Fourier operator. The matrix $\mathbf{B}(t)$ is proportional to
the quadrupolar matrix of Enceladus \citep[equation (43)]{RaR2017}.
\begin{equation}
V'(z,t) = -\frac{3}{2}\frac{G {\rm I}_\circ}{|z|^5} 
\, \, \, z^T\,\mathbf{B}(t)\, z\, .
\end{equation}
The perturbing quadrupolar potential can be decomposed into  a centrifugal term and a tidal
term and can be written as 
\begin{equation}
  W(z,t) = - \frac{1}{2} \, \, \, z^T\,\mathbf{F}(t)\, z\, , 
\end{equation}
where
\begin{equation}\label{fbody}
\mathbf{F}(t) =-{\boldsymbol{\Omega}}^{2}+\frac{1}{3}{\rm \!\ Tr\!\ }({\boldsymbol{\Omega}}^{2})\mathbb {I}+
\frac{3G m_{2}}{|q|^{5}}\left(q\otimes q-\frac{|q|^{2}}{3} \mathbb {I}\right)\,.
\end{equation}
Therefore, equation~(\ref{LoveVW}) can be rewritten as
\begin{equation}\label{Lovefourier}
3 \frac{G{\rm I}_\circ}{R^5}\,\hat{\mathbf{B}}(\omega) = k_2(\omega)\,\hat{\mathbf{F}}(\omega)\,.
\end{equation}
This equation, written in the frequency domain, is not suitable for numerical integrations. 
In the following section we provide equivalent equations in the time domain for two particular
rheologies. The method being very general, it can be applied to arbitrary rheologies.


\section{The rheologies of Andrade and Maxwell}\label{sec_rheology}

Linear visco-elastic rheologies are usually represented by spring-dashpot systems. Adding to this system a spring of elastic constant $\gamma$, which
represents gravity, and a mass $\mu$, which represents the inertia due to deformations,  we can construct a one dimensional oscillator
(see \cite{RaR2017} for details).
The oscillators associated with the Maxwell and with the Andrade rheologies are  shown in Figures  \ref{maxwell-osc} and \ref{andrade-osc},
respectively. The total displacement of the oscillator is denoted by $x$.
The equations for the deformation matrix $\mathbf{B}$ are introduced by means of the
{\bf Association Principle} ({\bf AP}) derived by \cite{RaR2017}\footnote{The equivalence of the {\bf AP} and the Correspondence Principle \citep{efr2012} is addressed in Section 4 of \cite{Correia2018}. The main difference of the Association Principle and the Correspondence Principle is that the first is formulated in the time domain, while the second is formulated in the frequency domain.}:
\begin{itemize}
\item[] ``The differential equation for $\mathbf{B}$ in the body reference frame $\mathbf{K}$
  is equal to the differential equation for the viscoelastic oscillator after replacing  $x$ by $\mathbf{B}$''.
\end{itemize}


\subsection{Maxwell rheology}


\begin{figure}[ptb]
\begin{center}
\begin{tikzpicture}[scale=0.9, transform shape]
\tikzstyle{spring}=[thick,decorate,decoration={zigzag,pre length=0.5cm,post length=0.5cm,segment length=6}]
\tikzstyle{damper}=[thick,decoration={markings,  
  mark connection node=dmp,
  mark=at position 0.5 with 
  {
    \node (dmp) [thick,inner sep=0pt,transform shape,rotate=-90,minimum width=15pt,minimum height=3pt,draw=none] {};
    \draw [thick] ($(dmp.north east)+(5pt,0)$) -- (dmp.south east) -- (dmp.south west) -- ($(dmp.north west)+(5pt,0)$);
    \draw [thick] ($(dmp.north)+(0,-5pt)$) -- ($(dmp.north)+(0,5pt)$);
  }
}, decorate]
\tikzstyle{ground}=[fill,pattern=north east lines,draw=none,minimum width=0.75cm,minimum height=0.3cm]

            \draw [thick] (0,0.7) -- (1,0.7);
            \draw [thick] (1,0) -- (1,1.4);
            \draw [-latex, thick] (1.5,1.4) -- (0.5,1.4) node[above] {$f-\lambda$};
            \draw [-latex, thick] (1,0) -- (0.5,0) node[below] {$\lambda$};
            \draw [spring] (1,0) -- node[above] {$\alpha_{0}$} (3,0);
            \draw [damper] (3,0) -- (4,0);
            \node at (3.6,0.5) {$\eta_{0}$};
            \draw [-latex,thick] (4,0) -- (5,0) node[below] {$\lambda$};
            \draw [spring] (1.5,1.4) -- node[above] {$\gamma$} (4,1.4);
            \draw [-latex, thick] (4,1.4) -- (5,1.4) node[above] {$f-\lambda$};
            \draw [thick] (4.5,0) -- (4.5,1.4);
            \draw [thick] (4.5,0.7) -- (5,0.7);
            \draw [latex-latex, thick] (1,-0.7) -- node[below] {$x$} (4.5,-0.7);

\node (M) at (5.27,0.7) [minimum width=0.5cm, minimum height=0.5cm, style={draw,outer sep=0pt,thick}] {$\mu$};
\draw [-latex, thick] (M.east) ++ (0cm,0) -- +(1cm,0) node[below] {$f(t)$};
\node (wall) at (-0.15,0.5) [ground, rotate=-90, minimum width=3cm] {};
\draw [thick] (wall.north east) -- (wall.north west);
      \end{tikzpicture}
\end{center}
\caption{Maxwell oscillator. The external force $f(t)$ splits into the force $\lambda(t)$ that acts upon the Maxwell array plus the force $f(t)-\lambda(t)$ that acts upon the $\gamma$ spring.}
\label{maxwell-osc}
\end{figure}
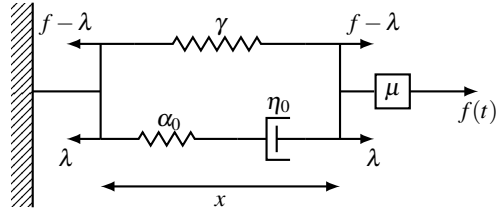


The oscillator associated with the Maxwell rheology is given  in Figure \ref{maxwell-osc}.
The equations of motion of the oscillator are
\begin{equation}
  \begin{split}
    \dot x&=u\\
\mu \dot u&=-\gamma x-\lambda+f\\
\alpha_0^{-1}\dot{\lambda}&=-\eta_{0}^{-1}\lambda+u,
\end{split}
\end{equation}
where $\mu$, $\gamma$, $\alpha_0$ and $\eta_0$ represent,  respectively: inertia of deformation, gravity, elasticity and viscosity.

Implementing the AP we obtain the following equations of motion for $\mathbf{B}$
\begin{equation}\label{maineq1}
\begin{split}
\dot {\mathbf{B}}&=\mathbf{U}\\
\mu\dot{\mathbf{U}}&=-\gamma \mathbf{B}-\boldsymbol{\Lambda}+\mathbf{F}\\
J_{0}\dot{\boldsymbol{\Lambda}}&=-\eta_{0}^{-1}\boldsymbol\Lambda+
\mathbf{U},
\end{split}
\end{equation}
where $\mathbf{F}(t)$ is the tidal forcing (\ref{fbody}), and $J_{0}=1/\alpha_{0}$.
With $\mu=0$ (see \cite{Correia2018}\footnote{In \cite{Correia2018} it was shown that deformation inertia is negligible if $\omega\ll\omega_{0}$, where $\omega_{0}=\sqrt{\gamma/\mu}$ is the natural frequency of oscillation of the system when damping is neglected and $\omega$ is the angular frequency of a harmonic tidal force.}) the equations of motion (\ref{maineq1}) can be written
\begin{equation}\label{eqmax0}
\dot {\mathbf{B}} = \frac{J_{0}}{1+\gamma J_{0}}\dot{\mathbf{F}}+\frac{\eta_{0}^{-1}}{1+\gamma J_{0}}(\mathbf{F}-\gamma\mathbf{B}),
\end{equation}
Then, applying the change of variables (\ref{inertfr}) together with the formula (\ref{timeder}), we get the
expression of the equation of motion in the inertial reference frame, namely,
\begin{equation}\label{eqmax}
\dot {\mathbf{b}} = [\boldsymbol {\omega},\mathbf{b}]+\frac{J_{0}}{1+\gamma J_{0}}(\dot{\mathbf{f}}-[\boldsymbol {\omega},\mathbf{f}])+\frac{\eta_{0}^{-1}}{1+\gamma J_{0}}(\mathbf{f}-\gamma\mathbf{b}),
\end{equation}

where
\begin{equation}\label{forcein}
\mathbf{f} =-{\boldsymbol {\omega}}^{2}+\frac{1}{3}{\rm \!\ Tr\!\ }({\boldsymbol {\omega}}^{2})\mathbb {I}+
\frac{3G m_2}{|q'|^5}\left(q'\otimes q'-\frac{|q'|^2}{3} \mathbb {I}\right).
\end{equation}

\subsection{Andrade rheology}

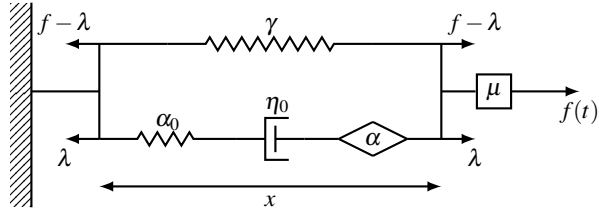
\begin{figure}[ptb]
\begin{center}
\begin{tikzpicture}[scale=0.9, transform shape]
\tikzstyle{spring}=[thick,decorate,decoration={zigzag,pre length=0.5cm,post length=0.5cm,segment length=6}]
\tikzstyle{damper}=[thick,decoration={markings,  
  mark connection node=dmp,
  mark=at position 0.5 with 
  {
    \node (dmp) [thick,inner sep=0pt,transform shape,rotate=-90,minimum width=15pt,minimum height=3pt,draw=none] {};
    \draw [thick] ($(dmp.north east)+(5pt,0)$) -- (dmp.south east) -- (dmp.south west) -- ($(dmp.north west)+(5pt,0)$);
    \draw [thick] ($(dmp.north)+(0,-5pt)$) -- ($(dmp.north)+(0,5pt)$);
  }
}, decorate]
\tikzstyle{ground}=[fill,pattern=north east lines,draw=none,minimum width=0.75cm,minimum height=0.3cm]

            \draw [thick] (0,0.7) -- (1,0.7);
            \draw [thick] (1,0) -- (1,1.4);
            \draw [-latex, thick] (2,1.4) -- (0.5,1.4) node[above] {$f-\lambda$};
            \draw [-latex, thick] (1,0) -- (0.5,0) node[below] {$\lambda$};
            \draw [spring] (1,0) -- node[above] {$\alpha_{0}$} (3,0);
            \draw [damper] (3,0) -- (4,0);
            \node at (3.6,0.5) {$\eta_{0}$};
            \draw [thick] (4,0) -- (4.5,0);
            \draw [thick] (4.5,0) -- (5,0.25);
            \draw [thick] (5,0.25) -- (5.5,0);
            \draw [thick] (5.5,0) -- (5,-0.25);
            \node at (5,0) {$\alpha$};
            \draw [thick] (5,-0.25) -- (4.5,0);
            \draw [-latex,thick] (5.5,0) -- (6.5,0) node[below] {$\lambda$};
            \draw [spring] (2,1.4) -- node[above] {$\gamma$} (5,1.4);
            \draw [-latex, thick] (5,1.4) -- (6.5,1.4) node[above] {$f-\lambda$};
            \draw [thick] (6,0) -- (6,1.4);
            \draw [thick] (6,0.7) -- (6.5,0.7);
            \draw [latex-latex, thick] (1,-0.7) -- node[below] {$x$} (6,-0.7);

\node (M) at (6.77,0.7) [minimum width=0.5cm, minimum height=0.5cm, style={draw,outer sep=0pt,thick}] {$\mu$};
\draw [-latex, thick] (M.east) ++ (0cm,0) -- +(1cm,0) node[below] {$f(t)$};
\node (wall) at (-0.15,0.5) [ground, rotate=-90, minimum width=3cm] {};
\draw [thick] (wall.north east) -- (wall.north west);
      \end{tikzpicture}
\end{center}
\caption{The Andrade  oscillator formulated from the Andrade model, $\alpha$ represents the so called spring-pot element. The external force $f(t)$ splits into the force $\lambda(t)$ that acts upon the Andrade array plus the force $f(t)-\lambda(t)$ that acts upon the $\gamma$ spring.}
\label{andrade-osc}
\end{figure}

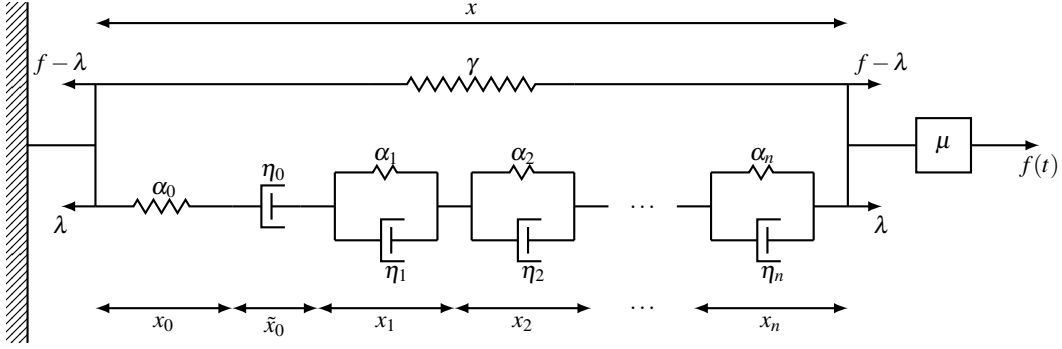
\begin{figure*}[ptb]
\begin{center}
\begin{tikzpicture}[scale=0.9, transform shape]
\tikzstyle{spring}=[thick,decorate,decoration={zigzag,pre length=0.5cm,post length=0.5cm,segment length=6}]
\tikzstyle{damper}=[thick,decoration=
{markings, mark connection node=dmp, mark=at position 0.5 with 
  {
    \node (dmp) [thick,inner sep=0pt,transform shape,rotate=-90,minimum width=15pt,minimum height=3pt,draw=none] {};
    \draw [thick] ($(dmp.north east)+(5pt,0)$) -- (dmp.south east) -- (dmp.south west) -- ($(dmp.north west)+(5pt,0)$);
    \draw [thick] ($(dmp.north)+(0,-5pt)$) -- ($(dmp.north)+(0,5pt)$);
  }
}, decorate]
\tikzstyle{ground}=[fill,pattern=north east lines,draw=none,minimum width=0.75cm,minimum height=0.3cm]

\node (wall) at (-0.15,0.5) [ground, rotate=-90, minimum width=5cm] {};
\draw [thick] (wall.north east) -- (wall.north west);

            \draw [latex-latex, thick] (1,2.7) -- node[above] {$x$} (12,2.7);

            \draw [thick] (0,0.9) -- (1,0.9);

            \draw [thick] (1,0) -- (1,1.8);

            \draw [-latex, thick] (5,1.8) -- (0.5,1.8) node[above] {$f-\lambda$};
            \draw [spring] (5,1.8) -- node[above] {$\gamma$} (8,1.8);
            \draw [-latex, thick] (8,1.8) -- (12.5,1.8) node[above] {$f-\lambda$};

            \draw [-latex, thick] (1,0) -- (0.5,0) node[below] {$\lambda$};
            \draw [spring] (1,0) -- node[above] {$\alpha_{0}$} (3,0);
            \draw [damper] (3,0) -- (4,0);
            \node at (3.6,0.5) {$\eta_{0}$};

            \draw [thick] (4,0) -- (4.5,0);

            \draw [thick] (4.5,-0.5) -- (4.5,0.5);
            \draw [spring] (4.5,0.5) -- node[above] {$\alpha_{1}$} (6,0.5);
            \draw [damper] (4.5,-0.5) -- (6,-0.5);
            \node at (5.4,-1.0) {$\eta_{1}$};
            \draw [thick] (6,-0.5) -- (6,0.5);

            \draw [thick] (6,0) -- (6.5,0);

            \draw [thick] (6.5,-0.5) -- (6.5,0.5);
            \draw [spring] (6.5,0.5) -- node[above] {$\alpha_{2}$} (8,0.5);
            \draw [damper] (6.5,-0.5) -- (8,-0.5);
            \node at (7.4,-1.0) {$\eta_{2}$};
            \draw [thick] (8,-0.5) -- (8,0.5);

            \draw [thick] (8,0) -- (8.5,0);
            \node at (9,0) {$\ldots$};
            \draw [thick] (9.5,0) -- (10,0);

            \draw [thick] (10,-0.5) -- (10,0.5);
            \draw [spring] (10,0.5) -- node[above] {$\alpha_{n}$} (11.5,0.5);
            \draw [damper] (10,-0.5) -- (11.5,-0.5);
            \node at (10.9,-1.0) {$\eta_{n}$};
            \draw [thick] (11.5,-0.5) -- (11.5,0.5);

            \draw [-latex,thick] (11.5,0) -- (12.5,0) node[below] {$\lambda$};

            \draw [thick] (12,0) -- (12,1.8);
            \draw [thick] (12,0.9) -- (13,0.9);

\node (M) at (13.4,0.9) [minimum width=0.8cm, minimum height=0.8cm, style={draw,outer sep=0pt,thick}] {$\mu$};

\draw [-latex, thick] (M.east) ++ (0cm,0) -- +(1cm,0) node[below] {$f(t)$};

            \draw [latex-latex, thick] (1,-1.5) -- node[below] {$x_{0}$} (3,-1.5);
            \draw [latex-latex, thick] (3,-1.5) -- node[below] {$\tilde{x}_{0}$} (4.25,-1.5);
            \draw [latex-latex, thick] (4.25,-1.5) -- node[below] {$x_{1}$} (6.25,-1.5);
            \draw [latex-latex, thick] (6.25,-1.5) -- node[below] {$x_{2}$} (8.25,-1.5);
            \node at (9,-1.5) {$\ldots$};
            \draw [latex-latex, thick] (9.75,-1.5) -- node[below] {$x_{n}$} (12,-1.5);

      \end{tikzpicture}
\end{center}
\caption{Extended Burgers oscillator associated with the Andrade model.}
\label{Andosc}
\end{figure*}

The oscillator associated with the Andrade rheology is given  in Figure \ref{andrade-osc}.
As will be shown in Section \ref{aprox} the Andrade element can be  approximated by an extended Burgers oscillator with $n$ Voigt elements given 
in Figure \ref{Andosc}, where $x=x_{0} + \tilde{x}_{0} + x_{1} + \ldots + x_{n}$ is the total displacement of the oscillator.
The equations of motion of the oscillator are
\begin{equation}
\begin{split}
\dot x&=u\\
\mu\dot u&=-\gamma x-\lambda+f\\
J_0\dot\lambda&=-\left(\sum_{r=0}^n\eta_r^{-1}\right)\lambda+u+
\sum_{r=1}^n\tau_r^{-1}x_r\label{osc}\\
\dot x_1&=-\tau_1^{-1}x_1+\eta_1^{-1}\lambda\\
\dot x_2&=-\tau_2^{-1}x_2+\eta_2^{-1}\lambda\\
\ldots\\
\dot x_n&=-\tau_n^{-1}x_n+\eta_n^{-1}\lambda,
\end{split}
\end{equation}
where $\eta_{r}$ is the viscosity of the $r$-th dashpot, $\tau_{r}=\eta_{r}/\alpha_{r}$ is the relaxation time of the $r$-th
Voigt element, $r=1,...,n$, and we used the relation
\[
\dot{x}_{0}+\dot{\tilde{x}}_0=\dot{x} - (\dot{x}_{1} + \ldots + \dot{x}_{n})=
\dot\lambda/\alpha_0+\lambda/\eta_0.
\]

By means of the AP we find the equations
for $\mathbf B$, $\boldsymbol\Lambda$,
and $\mathbf{B}_r$, $r=1,\ldots, n$  ($\boldsymbol\Lambda$ and $\mathbf{B}_r$
are symmetric traceless matrices):
\begin{equation}\label{maineq2}
\begin{split}
\dot {\mathbf{B}}&=\mathbf{U}\\
\mu\dot{\mathbf{U}}&=-\gamma \mathbf{B}-\boldsymbol\Lambda+\mathbf{F}\\
J_0\dot{\boldsymbol\Lambda}&=-(\sum_{r=0}^n\eta_r^{-1})\boldsymbol\Lambda+
\mathbf{U}+\sum_{r=1}^n\tau_r^{-1}\mathbf{B}_r\\
\dot {\mathbf{B}}_1&=-\tau_1^{-1}\mathbf{B}_1+\eta_1^{-1}\boldsymbol\Lambda\\
\dot {\mathbf{B}}_2&=-\tau_2^{-1}\mathbf{B}_2+\eta_2^{-1}\boldsymbol\Lambda\\
\ldots\\
\dot {\mathbf{B}}_n&=-\tau_n^{-1}\mathbf{B}_n+\eta_n^{-1}\boldsymbol\Lambda\, ,
\end{split}
\end{equation}
where $\mathbf{F}$ is the tidal forcing in equation (\ref{fbody}). In the inertial reference frame and   with $\mu=0$ these equations become
\begin{equation}\label{eq11}
\begin{split}
\dot {\mathbf{b}}&= [\boldsymbol {\omega},\mathbf{b}]+\frac{J_{0}}{1+\gamma J_{0}}(\dot{\mathbf{f}}-[\boldsymbol {\omega},\mathbf{f}])\\
&+\frac{\sum_{r=0}^n\eta_r^{-1}}{1+\gamma J_{0}}(\mathbf{f}-\gamma\mathbf{b})-\frac{\sum_{r=1}^n\tau_r^{-1}\mathbf{b}_r}{1+\gamma J_{0}}\\
\dot {\mathbf{b}}_1&= [\boldsymbol {\omega},\mathbf{b}_{1}]-\tau_1^{-1}\mathbf{b}_1+\eta_1^{-1}(\mathbf{f}-\gamma\mathbf{b})\\
\dot {\mathbf{b}}_2&= [\boldsymbol {\omega},\mathbf{b}_{2}]-\tau_2^{-1}\mathbf{b}_2+\eta_2^{-1}(\mathbf{f}-\gamma\mathbf{b})\\
\ldots\\
\dot {\mathbf{b}}_n&= [\boldsymbol {\omega},\mathbf{b}_{n}]-\tau_n^{-1}\mathbf{b}_n+\eta_n^{-1}(\mathbf{f}-\gamma\mathbf{b}),
\end{split}
\end{equation}
with $\mathbf{f}$ from equation (\ref{forcein}).


\section{An approximation to the Andrade rheology}\label{aprox}

In this Section we show that the spring-pot element in the spring-dashpot representation of the
Andrade rheology, Fig. \ref{andrade-osc}, can be replaced  by a continuum of Voigt elements in series and further approximated
by a finite set of $n$ Voigt elements as in Fig. \ref{Andosc}. In the particular example of this paper
we show that $n$ can be taken equal to six. The same procedure can be done in any other situation to determine
a minimum number of Voigt elements $n$ that can be used in the approximation.


\subsection{The Andrade model}
The \cite{and1910} model is represented by three mechanical elements combined in series: a spring, a dashpot and a spring-pot (fig. \ref{andrade-osc}). The  creep function of the model is
\begin{equation}
J(t)=\Big[J_{0}+\frac{t}{\eta_{0}}+At^{\alpha}\Big]H(t),
\end{equation}
where $H$ is the Heaviside function, and $A>0$ and $0<\alpha<1$ are physical parameters.
In order to avoid a fractional dimension of time, \cite{efr2012} replaces the parameter $A$ by $\tau_{A}$,
the so called ``Andrade time'' (with dimension of time):
\begin{equation}
A = J_{0}\tau_{A}^{-\alpha} = \alpha_{0}^{-1}\tau_{A}^{-\alpha}.
\end{equation}
In terms of the  Andrade and Maxwell $\tau_{M}=\eta_{0}/\alpha_{0}$ times the creep  function becomes
\begin{equation}
J(t)=J_{0}\Big[1+\frac{t}{\tau_{M}}+\Big(\frac{t}{\tau_{A}}\Big)^{\alpha}\Big]H(t).
\end{equation}
This creep function induces the following  complex  compliance in the frequency domain
\begin{equation}\label{Andradecc}
\hat{J}(\omega) =J_{0}[1+(i\omega\tau_{A})^{-\alpha}\Gamma(1+\alpha)-i(\omega\tau_{M})^{-1}].
\end{equation}


\subsection{The representation of the Andrade model by an extended Burgers model}
The creep function of a finite array of $n$ Voigt elements is \citep{bla2016}
\begin{equation}
J(t)=\sum_{r=1}^{n}J_{r}[1-\exp(-t/\tau_{r})]H(t),
\end{equation}
where $J_{r}$ and $\tau_{r}=\eta_{r}J_{r}$ are the compliance and the relaxation time of the $r^{th}$ Voigt element, respectively.
The complex compliance of this finite array is
\begin{equation}\label{cc}
i\omega\hat{J}(\omega)=\sum_{r=1}^{n}\frac{J_{r}}{1+i\omega\tau_{r}}.
\end{equation}
The continuum limit of this array is well defined as $n\to\infty$ provided that  $\sum_{r=1}^{n}J_r$ converges.
In the limit $n\to\infty$, the creep function becomes
\begin{equation}
J(t)=H(t)\int_{0}^{\infty}j(\tau)[1-\exp(-t/\tau)]d\tau.
\end{equation}
In order to obtain the continuous distribution of compliances associated with the Andrade model it is necessary to solve the equation
\begin{equation}
At^{\alpha}=\int_{0}^{\infty}j({\tau})[1-\exp(-t/\tau)]d\tau
\end{equation}
for $j(\tau)$. The solution to the equation is
\begin{equation}
j(\tau)=\frac{A\alpha}{\Gamma(1-\alpha)}\tau^{-1+\alpha}, \qquad 0<\alpha<1.
\end{equation}
Using the last two equations the Fourier transform of $At^{\alpha}H(t)$ can be written as
\begin{equation}
\int_{-\infty}^{\infty}At^{\alpha}H(t)e^{-i\omega t}dt=\frac{A\alpha}{\Gamma(1-\alpha)}\int_{0}^{\infty}\frac{\tau^{-1+\alpha}}{i\omega-\tau\omega^{2}}d\tau.
\end{equation}
Then the  creep function becomes
\begin{equation}
J(t)=\Bigg[J_{0}+\frac{t}{\eta_{0}}+\frac{A\alpha}{\Gamma(1-\alpha)}\int_{0}^{\infty}\tau^{-1+\alpha}[1-\exp(\frac{-t}{\tau})]d\tau\Bigg]H(t)
\end{equation}
and the complex compliance becomes
\begin{equation}\label{cont}
\hat{J}(\omega)=J_{0}+(i\eta_{0}\omega)^{-1}+\frac{A\alpha}{\Gamma(1-\alpha)}\int_{0}^{\infty}\frac{\tau^{-1+\alpha}}{1+i\tau\omega}d\tau.
\end{equation}
The main idea in the discretisation of the Andrade model is to approximate the integral in the last term of equation (\ref{cont})
by a  finite sum, equation (\ref{cc}),  that represents the  complex compliance of an extended
Burgers model, namely
\begin{equation}\label{55}
\frac{A\alpha}{\Gamma(1-\alpha)}\int_{0}^{\infty}\frac{\tau^{-1+\alpha}}{1+i\tau\omega}d\tau\approx\sum_{r=1}^{n}\frac{J_{r}}{1+i\omega\tau_{r}}.
\end{equation} 
One way to do this is described in \cite{Birk2010}. The compliance $J_{r}$  and the relaxation time $\tau_{r}$ of the Voigt elements in equation (\ref{55}) for $r=1,\dots,n$ are given by
\begin{equation}\label{jrtr}
J_{r}=\frac{8A\alpha}{\Gamma(1-\alpha)}\frac{\bar{\lambda}_{r}}{(1-\bar{q}_{r})^{4}}, \qquad \tau_{r}=\frac{(1+\bar{q}_{r})^{4}}{(1-\bar{q}_{r})^{4}},
\end{equation}
where $\bar{\lambda}_{r}$ and $\bar{q}_{r}$ denote the weights and abscissas, respectively, of the $n$-point Gauss-Jacobi quadrature rule, with weighting function $(1-\bar{q})^{1-2\alpha}(1+\bar{q})^{-1+4\alpha}$.


\subsubsection{The number of Voigt elements}\label{number}


\begin{figure}[ptb]
\centering
\includegraphics[scale=0.35]{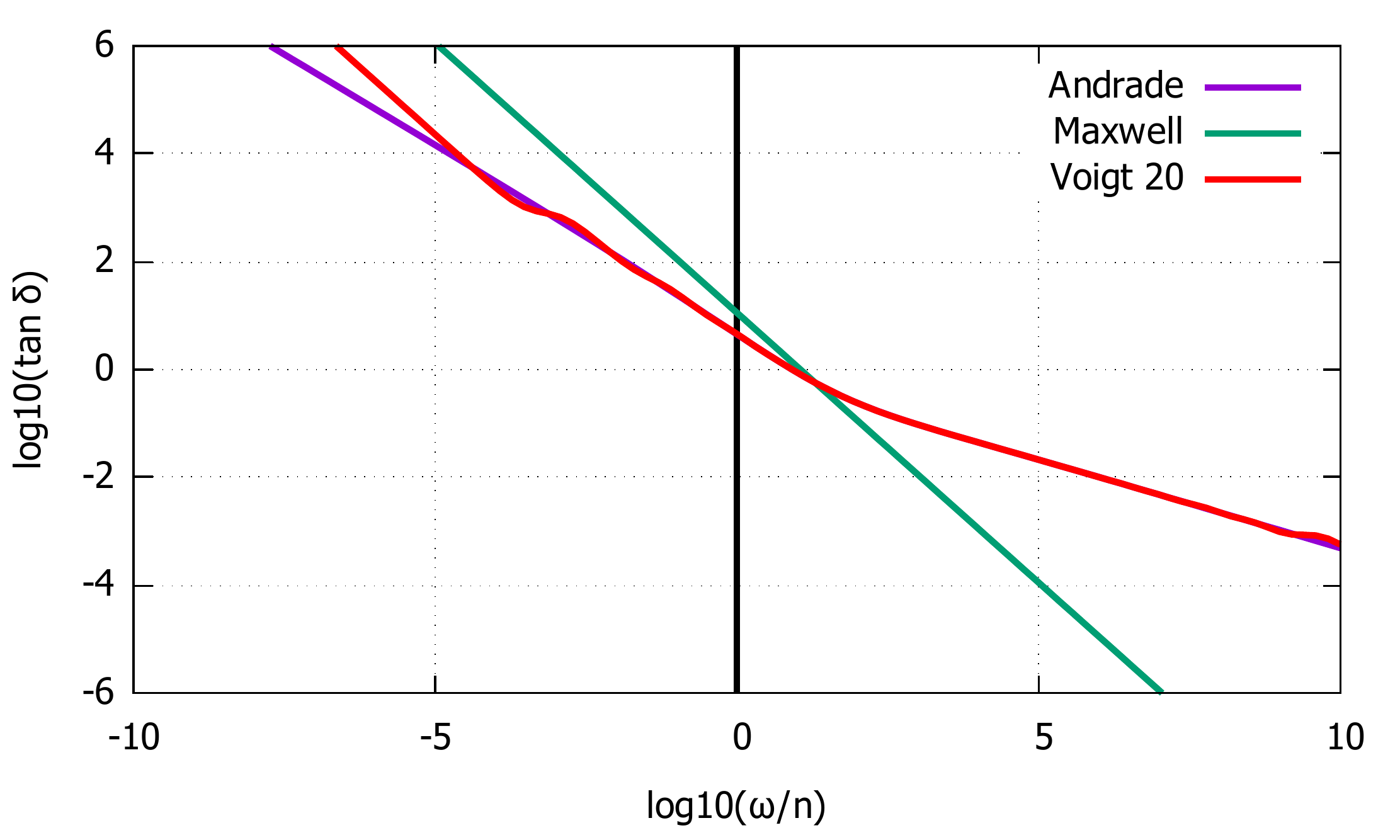} \\
\includegraphics[scale=0.35]{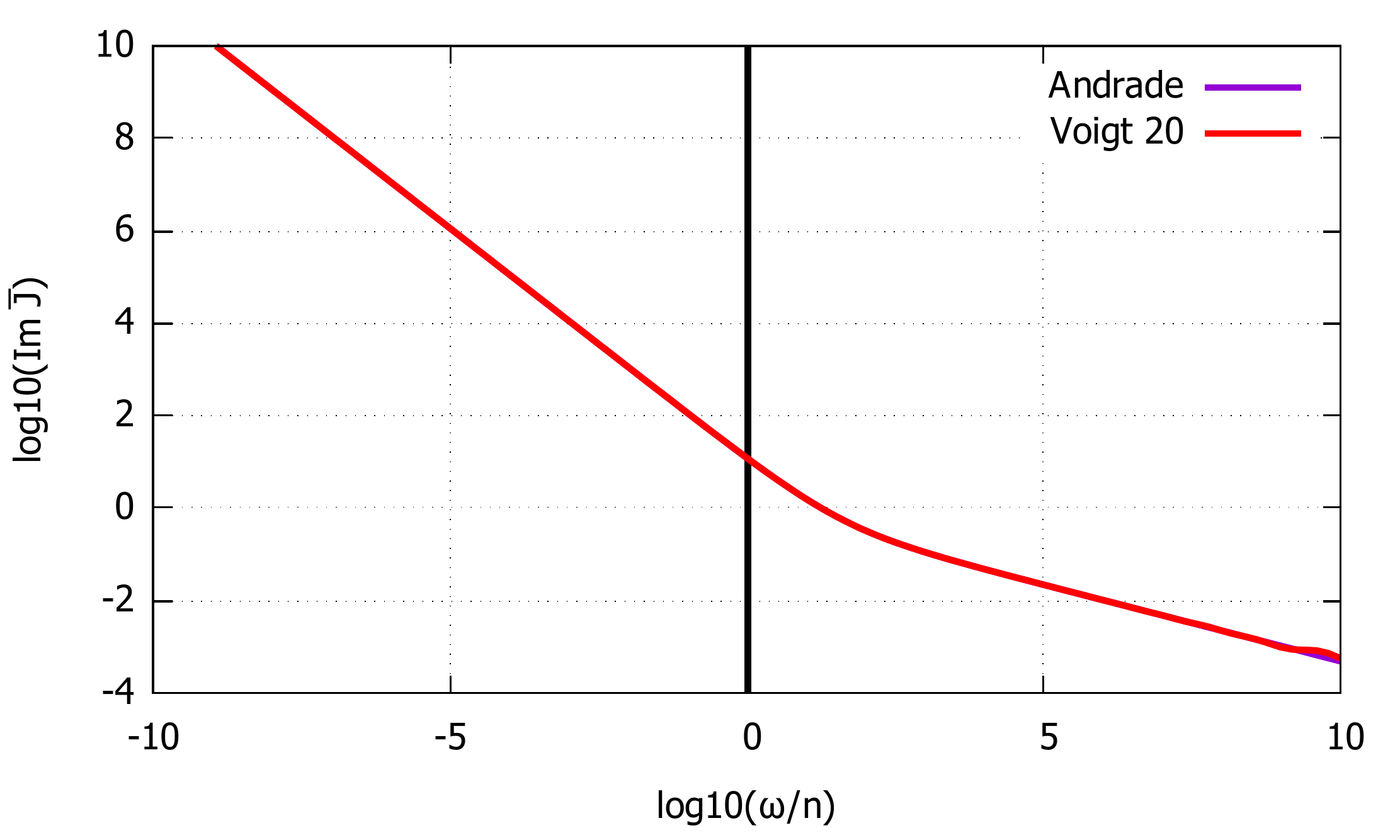}
\caption{Approximation of the Andrade rheology by an extended Burgers model with 20 Voigt elements. The phase lag (top) and the compliance imaginary part (bottom) are represented as a function of the excitation frequency $\omega$ in units of the orbital frequency $n$. The black vertical line marks the value of Enceladus mean rotation rate.\label{fig:andrade_20}}

\end{figure}


\begin{figure}[ptb]
\centering
\includegraphics[scale=0.35]{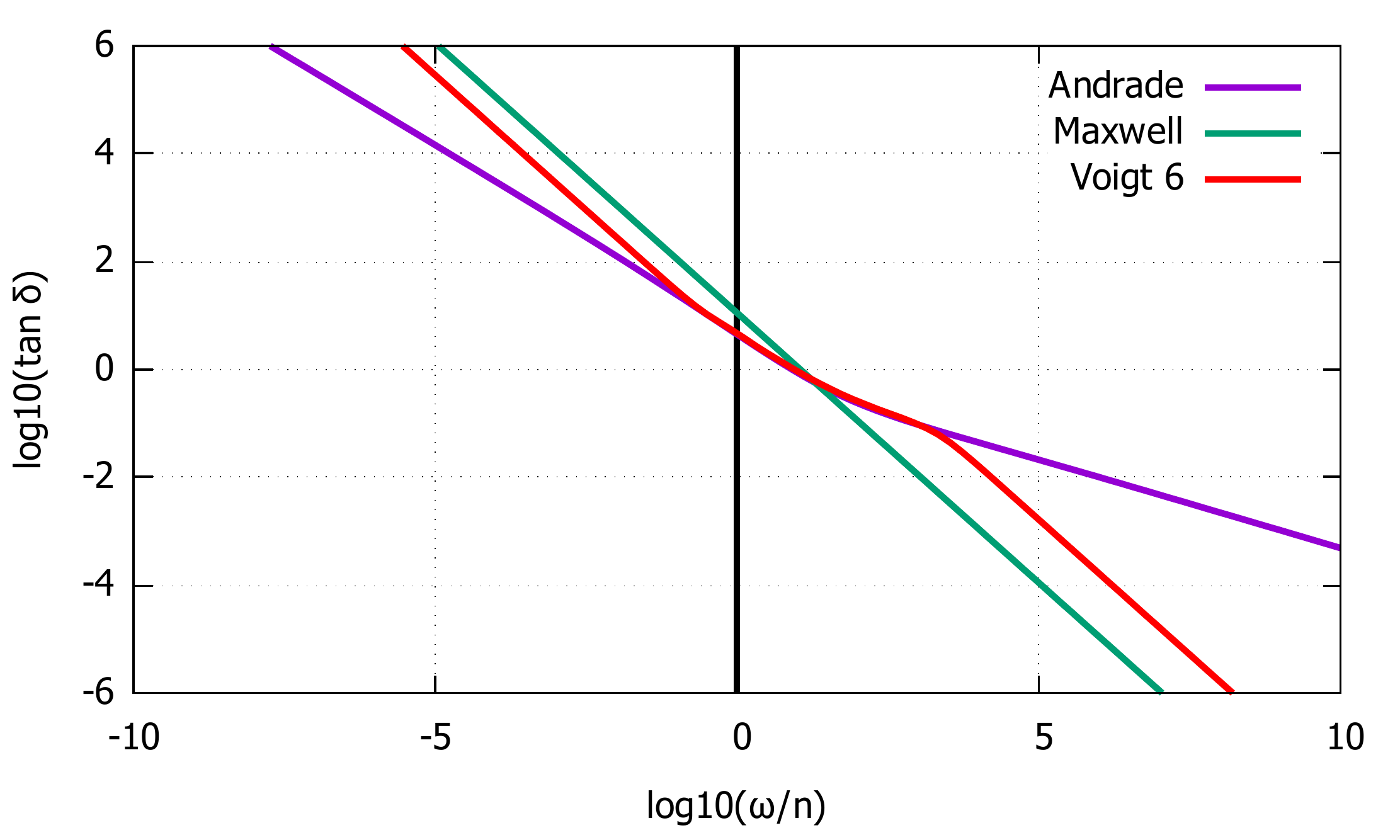} \\
\includegraphics[scale=0.35]{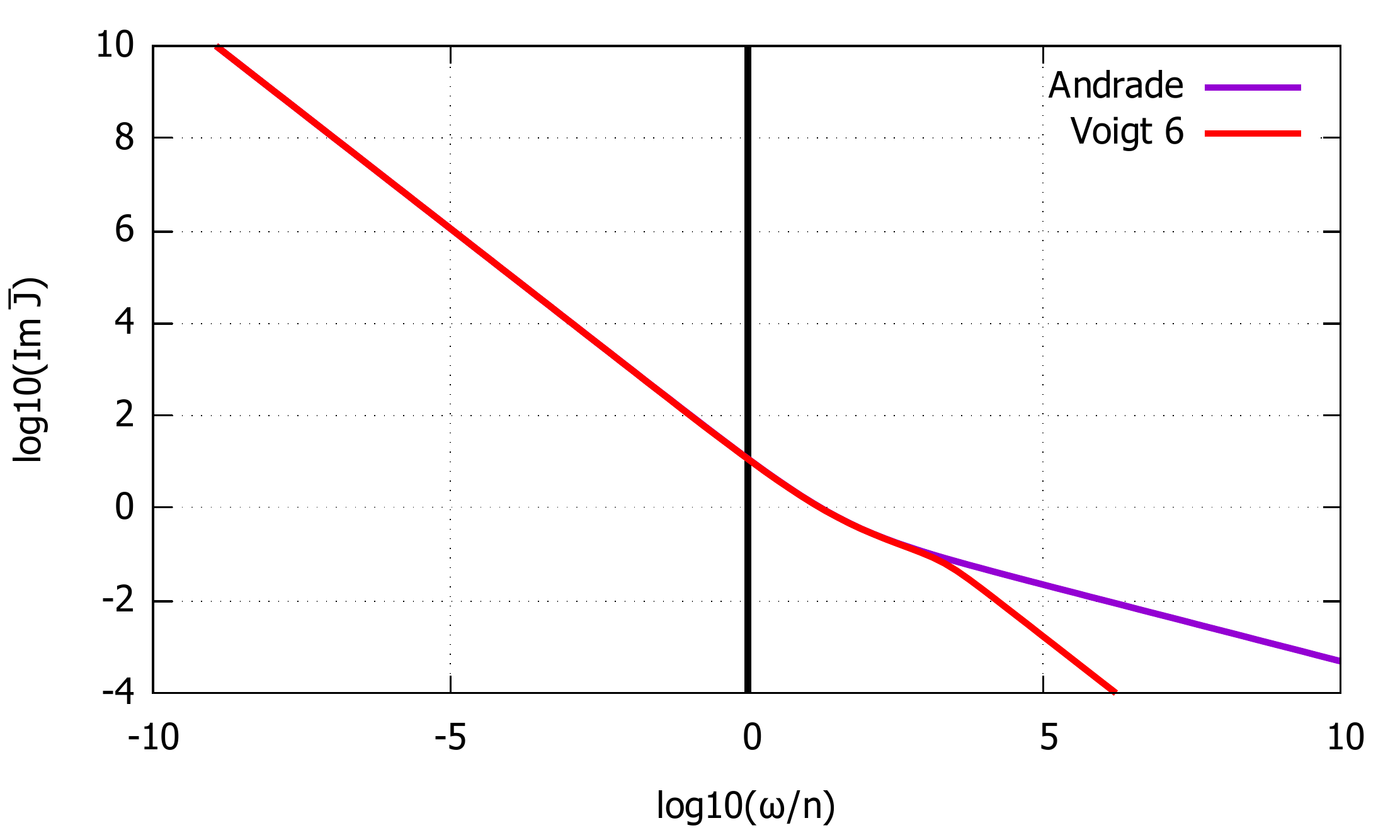}
\caption{Same as Fig.~\ref{fig:andrade_20} but for an extended Burgers model made of 6 Voigt elements only.\label{fig:andrade_6}}
\end{figure}

In the following, we fix   the parameters of the Andrade and Maxwell models as  $\alpha = 0.33$, $J_0=1/\alpha_0=10^{10}\,\mathrm{Pa}$,
$\eta_{0}=0.17\times 10^{14}\,\mathrm{Pa}\cdot\mathrm{s}$, and $\tau_M=\tau_{A} = 1700 s$ (see next section).
In this section these values and the  dominant  tidal forcing frequency upon Enceladus 
are used as prototypes. The goal is to show how to estimate a number of Voigt elements in an extended Burgers model that
guarantees a good approximation for the Andrade rheology in the frequency range of the tidal forcing.

The complex compliance of an extended Burgers model that approximates the Andrade model is
\begin{equation}\label{compl}
\hat{J}(\omega) = J_{0} + (i\eta_{0}\omega)^{-1} + \sum_{i=1}^{n}J_{i}\Big(\frac{1}{1+i\omega\tau_{i}}\Big),
\end{equation}
where $J_i$ and $\tau_i$ are given in equation (\ref{jrtr}).  
The phase lag associated with this complex compliance is 
\begin{equation}
\tan\delta(\omega) = -\frac{\Im[\hat{J}(\omega)]}{\Re[\hat{J}(\omega)]} = \frac{\frac{1}{\tau_{m}\omega}+\sum_{i=1}^{n}\frac{J_{i}}{J_{0}}\frac{\omega\tau_{i}}{1+(\omega\tau_{i})^{2}}}{1+\sum_{i=1}^{n}\frac{J_{i}}{J_{0}}\frac{1}{1+(\omega\tau_{i})^{2}}}.
\end{equation}
In Figure \ref{fig:andrade_20}, we show the phase lag and the imaginary part of the complex
compliance deduced from the Andrade model on the one hand and from the extended Burgers model with
20 Voigt elements on the other as a function of the frequency.  In Figure \ref{fig:andrade_6} we
show the same comparisons replacing the extended Burgers model with $20$  Voigt elements by another
with  $6$ elements.  As can be seen from these figures, the use of an extended Burgers rheology with
$6$ Voigt elements is sufficient to approximate the Andrade rheology within the range
$0.1\le\omega/n\le 2000$. In this case, this is enough  because the main fundamental frequency of
tidal forcing of Enceladus by Saturn actually is $n=5.308\times 10^{-5}\,\mathrm{rad} \cdot
\mathrm{s}^{-1}$ (the orbital frequency of Enceladus) therefore all harmonics of order $k$ with
$1\le k\le 2000$ are within the region where the approximation is good.


\section{Expected forced librations}\label{sec_libration}  

Forced librations may be an important source of energy dissipation in Enceladus (see
\cite{EFR2018328, efr2018}).  In this section  we derive  analytical expressions for the  amplitude
and dissipation power of forced librations under the hypotheses that the amplitude is small and the
spin axis is perpendicular to the orbital plane. This calculation will allow us to validate the
numerical results presented  in Section \ref{sec_integration}.

\subsection{Amplitude of libration}

The satellite  frame $\mathbf{K}$ and  the inertial frame $\kappa$, which   are defined by the
orthonormal vectors $(E_1,E_2,E_3)$ and  $(e_1,e_2,e_3)$, respectively, are chosen to have   $E_3$
and $e_3$   aligned with the satellite spin axis.  The angle from $e_1$ to $E_1$ is denoted by
$\theta$. If $M$ is the mean anomaly of  the satellite, then $\theta=M+\zeta$, where $\zeta$ is the
libration angle.  The angular velocity is $\dot\theta=\Omega_3=n+\dot\zeta$, where $n$ is the mean
rotation rate that is equal to the mean motion\footnote{At low viscosity the rotation becomes
asynchronous with an average angular velocity given by $\dot\theta \approx n(1 + 6e^2)$
\citep[e.g.,][]{Folonier2018}. Here this excess of rotation speed is neglected because calculations
are limit to first order in eccentricity.}.  

We now focus on the first equation in
(\ref{eqmot}), where $\mathbf{B}$ is determined by the generic equation (\ref{Lovefourier}) that with the notation
$K_2(\omega)=\frac{R^5}{3G{\rm I}_\circ} k_2(\omega)$ becomes
\begin{equation}\label{Love}
\hat {\mathbf{B}}(\omega) = K_2(\omega) \hat{\mathbf{F}}(\omega)\;.
\end{equation}
As it was shown  by \cite{Correia2018} (Section 4, in particular equation (31)), 
the Association Principle in the frequency domain implies  
\begin{equation}
  K_2(\omega)=\frac{1}{\gamma+\hat{J}^{\,-1}(\omega)-\mu\omega^2},
\end{equation}
an expression that holds for arbitrary rheology. If the inertia of deformation is neglected ($\mu=0$), as assumed in this paper, then\footnote{This expression is equivalent to
$K_2(n)=\frac{R^5}{3GI_o}k_2(n)$, where \[k_2(n)=\frac{3}{2}\frac{1}{1+\frac{57}{8\pi}\frac{1}{G\rho^2R^2}\frac{1}{\hat{J}}}\] used by \cite{Efroimsky2015} (Eq.36). Detailed comparison of two notations can be found in \citep[section 4]{Correia2018}. }
\begin{equation}\label{K2}
  K_2(\omega)=\frac{1}{\gamma+\hat{J}^{\,-1}(\omega)}.
\end{equation}

 For instance, for a Maxwell rheology the complex compliance is $\hat{J}(\omega)=\alpha_0^{-1}+(i\omega\eta_0)^{-1}$
and equation (\ref{K2}) implies
\begin{equation}\label{K2Maxwell}
K_2(\omega) = \frac{1}{\gamma} \frac{1+i\omega\tau_M}{1+i\omega(\tau_M+\tau_F)},
\end{equation}
where
\[
\tau_M = \eta_0/\alpha_0 \quad \mathrm{and} \quad \tau_F = \eta_0/\gamma
\]
are the Maxwell time and the fluid time (only due to gravity and viscosity), respectively.

The forcing term $\mathbf{F}$ is given in equation (\ref{fbody}). It contains a term that
represents the centrifugal stress,
$-{\boldsymbol{\Omega}}^{2}+\frac{1}{3}{\rm \!\ Tr\!\ }({\boldsymbol{\Omega}}^{2})\mathbb {I} $, and
another that represents the tidal stress. The centrifugal part is given by a diagonal matrix with equal
elements in the first and second lines. 
This term has no
effect on the libration and will be neglected. So,  
\begin{equation}\label{fbody2}
\mathbf{F}(t) =\frac{3G m_{2}}{|q|^{5}}\left(q\otimes q-\frac{|q|^{2}}{3} \mathbb {I}\right),
\end{equation}
which is the tidal stress.
The position of Saturn in $\mathbf{K}$ is given by
\begin{equation}\label{q}
q=r\begin{pmatrix}
\cos (f-\theta) \\
\sin (f-\theta)\\
0
\end{pmatrix},
\end{equation}
where $f$ is the true anomaly of Enceladus and  $r$ is the distance from Enceladus to  Saturn.
The first equation in (\ref{eqmot}) can be simplified. 
Since the motion is planar, $B_{13}=B_{23}=0$ and $[\mathbf{B},\boldsymbol{\Omega}^2]=0$.
Moreover, $\left[\mathbb {I},\mathbf{B}\right]=0$ implies that   the last term in right hand side of equation
the first equation in (\ref{eqmot}) can be written as $\left[\mathbf{F},\mathbf{B}\right]$. So, the first equation in (\ref{eqmot}) becomes
\begin{equation}\label{eqom2}
 \begin{split} \dot{\boldsymbol{\Omega}}+\dot{\mathbf{B}}\boldsymbol{\Omega}+\boldsymbol{\Omega}\dot{\mathbf{B}}
    +\mathbf{B}\dot{\boldsymbol{\Omega}}+\dot{\boldsymbol{\Omega}} \mathbf{B}
   &=\left[\mathbf{F},\mathbf{B}\right]\quad\text{or}\\
   \dot\Omega_3&=\frac{(B_{11}-B_{22})F_{12}+B_{12}(F_{22}-F_{11})+\Omega_3\dot{B}_{33}}{1-B_{33}} \, .
   \end{split}
\end{equation}
Equations (\ref{Love}),  (\ref{fbody2}), (\ref{q}),
and (\ref{eqom2}) determine the librations of Enceladus.

If the eccentricity  of the orbit $e$ is zero then there is a trivial  solution $\zeta(t)=0$.
If $e>0$ is small then there exists a nontrivial solution $\zeta$ that
has an amplitude proportional to  $e$ and a period equal to the orbital period.
In order to determine this solution up to order $e$ we use the following well known relations
\citep{yoder1995astrometric}
(terms of order $e^2$ will be neglected in  the following  equations):
\begin{equation}\label{erel}
  \frac{a}{r}=1+e\cos M,\quad f-M=2e\sin M \, ,
\end{equation}
where $a$ is the semi-major axis.
Using these relations, equations (\ref{fbody2}) and (\ref{q}),
and $f-\theta=f-M-(\theta-M)=2e\sin M-\zeta$ we can write the non trivial components of
$\mathbf{F}$ as 
\begin{equation}\label{Fij}
  \begin{split}
    F_{11}&=c_1(2/3+2e\cos M) 
    \\
    F_{12}&=c_1(2e\sin M-\zeta)=F_{21}\\
    F_{22}&=-c_1(1/3+e\cos M)=F_{33}\\
    c_1&=\frac{3G m_{2}}{a^3} \, .
  \end{split}
\end{equation}

Notice that $\mathbf{F}$  is the sum of a constant term plus an oscillating term with period $2\pi/n$.
So, $\mathbf{B}$ must be of the same form. Solving equation (\ref{Love})
we obtain that  the constant elements of $\mathbf{B}$ are 
\begin{equation}\label{ovB}
  \overline B_{11}=K_2(0)\,c_1\frac{2}{3},\qquad
  \overline B_{22}=-K_2(0)\,c_1\frac{1}{3}=\overline B_{33}\;.
\end{equation}
It is convenient to write the oscillating part of $\mathbf{B}$ and other quantities in complex
notation
\begin{equation}\label{oscparts}
  \begin{split}
   \tilde  B_{jk} &= \hat B_{jk} \exp(iM) + \mathrm{c.c.}\;,\\
   \tilde  F_{jk} &= \hat F_{jk} \exp(iM) + \mathrm{c.c.}\;,\\
   \tilde  \zeta      &= \hat \zeta      \exp(iM) + \mathrm{c.c.}\;,
\end{split}
\end{equation}
where $\mathrm{c.c.}$ stands for complex conjugate.
With these definitions,   equation (\ref{Love}) and $\dot M=n$ imply
\begin{equation}\label{Bij}
  \hat B_{jk} = K_2(n) \hat F_{jk}\ .
\end{equation}

  Equation (\ref{Fij}) shows that $\hat F_{12}$ is of the order of $e$ and so is $\hat B_{12}$,
  due to  equation (\ref{Bij}). Therefore up to the order of $e$ equation (\ref{eqom2}) becomes
  \begin{equation}\label{eqom3}
  \begin{split}  
-n^2\hat \zeta(1-\overline B_{33})&=in^2\hat B_{33}+(\overline B_{11}-\overline B_{22})\hat F_{12}+\hat B_{12}(\overline F_{22}-\overline F_{11}) \implies  \\
-n^2\hat \zeta(1+\frac{1}{3}c_1K_2(0))&=in^2K_2(n)\hat F_{33}+c_1\left[K_2(0) \hat F_{12}-\hat B_{12}\right] \implies\\
-n^2\hat \zeta(1+\frac{1}{3}c_1K_2(0))&=in^2K_2(n)\hat F_{33}+c_1\left[K_2(0) - K_2(n)\right] \hat F_{12} \,.
\end{split}
\end{equation}
Equation (\ref{Fij}) implies that $\hat F_{12}=-c_1(ie+\hat\zeta)$ and $\hat F_{33}=-c_1e/2$. This and equation  (\ref{eqom3}) imply
that the amplitude of libration $\beta$ is given by
\begin{equation}\label{zetae}
\beta=2|\hat \zeta|,\quad \hat \zeta = \frac{\frac{1}{2}ic_1K_2(n)-i(c_1/n)^2\left[K_2(n)-K_2(0)\right]}{1+\frac{1}{3}c_1K_2(0)+(c_1/n)^2\left[K_2(n)-K_2(0)\right]}\, e.
\end{equation}
Notice that the factor two accounts for the two exponential terms in the expression of $\tilde\zeta$ (\ref{oscparts})
with frequencies $n$ and $-n$, respectively.

\subsection{Amplitude of forced libration: comparison with other results in the literature}\label{compampl}

In this paper we do not a priori  assume
that Enceladus has a triaxial shape (the same is done in \cite{Folonier2018}).
The eventual triaxiality  is a consequence of the satellite motion and its  rheology. So,
before comparing our results with others in the literature,  it is necessary to analyse the mean moments of inertia of Enceladus given
in equation (\ref{ovB}). The following results hold for any rheology such that $J^{-1}(0)=0$ and whenever the inertia of deformation is neglected,
$\mu=0$.
 The first hypothesis is  valid for the
Maxwell and Andrade rheologies and for any other for which the static shape of the body is determined solely by gravity.

If $\overline A<\overline B<\overline C$ denote the three principal moments of inertia of Enceladus then the definition of $\mathbf{B}$ implies
\begin{equation}\label{B11}
  \frac{\overline B-\overline A}{\overline C}=\frac{\overline B_{11}-\overline B_{22}}{1-\overline B_{33}}\approx
  \overline B_{11}-\overline B_{22}\, ,
  \end{equation}
  where we used that $B_{kj}\ll 1$ for all $k,j$. The gravitational modulus  $\gamma$ is related to the zero secular Love number by \citep[equation (14)]{RaR2017}
  \begin{equation}\label{1gamma}
    \frac{1}{\gamma}=\frac{R^5}{3G{\rm I}_\circ} k_2(0)=K_2(0) \,.
  \end{equation}  
  The combination of  equations (\ref{B11}), (\ref{1gamma}),  and
   (\ref{ovB})
  gives
  \begin{equation}\label{ABC}
    \frac{\overline B-\overline A}{\overline C}\approx \frac{c_1}{\gamma}\, .
  \end{equation}
  An algebraic manipulation shows that the term in the numerator of
  equation (\ref{zetae}) can be written as
\begin{equation}\label{c1n}
  \frac{1}{3}c_1K_2(0)+(c_1/n)^2\left[K_2(n)-K_2(0)\right]=\frac{1}{3}\frac{c_1}{\gamma}+3\frac{c_1}{\gamma}\frac{m_2}{m_1+m_2}\left[\frac{k_2(n)}{k_2(0)}-1\right] \,.
\end{equation}
In \cite{Correia2018} (Section 5 Proposition 1), under the hypothesis $\mu=0$,
it is shown that $|k_2(n)/k_2(0)|\le 1$ for any $n\ge 0$. This implies that
the quantity in equation (\ref{c1n}) is of the order of $c_1/\gamma$ that is much smaller than one  due to equation (\ref{ABC}).
Therefore equation (\ref{zetae}) implies
\begin{equation}\label{zetae2}
  \hat \zeta \approx \bigg(\frac{1}{2}ic_1K_2(n)-i(c_1/n)^2\left[K_2(n)-K_2(0)\right]\bigg)\, e
\end{equation}
and, using equations  (\ref{ABC}), (\ref{c1n}) and $K_2(n)=\frac{k_2(n)}{k_2(0)}\frac{1}{\gamma}$ 
\begin{equation}\label{zetaapprox}
 \hat \zeta \approx i\,\frac{1}{2}\,e\,\frac{\overline B-\overline A}{\overline C}\frac{k_2(n)}{k_2(0)}+
 i\,3\,  e\,   \frac{\overline B-\overline A}{\overline C}\frac{m_2}{m_1+m_2}\left[1-\frac{k_2(n)}{k_2(0)}\right] \,.
 \end{equation}

In the limit $m_1\ll m_2$, the amplitude of libration $\beta=2|\hat\zeta|$ of a deformable body in hydrostatic equilibrium is thus
\begin{equation}\label{betaapprox}
 \beta\approx
 6\,  e\,   \frac{\overline B-\overline A}{\overline C} \left|1-\frac{5}{6}\frac{k_2(n)}{k_2(0)}\right| \,,
 \end{equation}
which coincides with the result obtained by \cite{VANHOOLST2013299}. This expression is compared with other results in literature by \cite{Noyelles2017}.

 The classical formula for the amplitude of forced libration of a rigid triaxial body under the same circumstances is 
 (\cite{danby1962} also \cite{EFR2018328} equation (13) or \cite{THOMAS201637} equation (1))
 \begin{equation}\label{zetaefro}
 \beta=2|\hat \zeta| \approx
 6\, e\,  \frac{\overline B-\overline A}{\overline C} \,.
\end{equation}
So our formula (\ref{betaapprox})  shows that
the amplitude of libration of a deformable body under hydrostatic equilibrium must be smaller than that of a rigid body with the
same principal moments of inertia $\overline A, \overline B, \overline C$, by a  factor
$|1-5k_2(n)/6k_2(0)|$.  The amplitudes are the same only if the complex number $k_2(n)$ is zero.

 The case  $k_2(n)=k_2(0)$, which holds for a body made of a perfect fluid  under self-gravity
 (no dissipation of energy)
 and  no inertia of deformation ($\mu=0$), has an interesting interpretation  that reveals
 the essence of a ``Tisserand frame''. In this case the satellite
 do have a tidal deformation,  which is easily seen from equation (\ref{Bij}),  with a  principal axis    
 aligned with the position axis of the planet. No torque acts upon the satellite and the satellite angular momentum remains constant. Nevertheless, the amplitude of the tidal deformation depends on the  distance from the satellite to the planet. As this distance varies the moment of inertia of the satellite also varies and conservation of angular momentum requires a variation of  angular velocity. Equation (\ref{betaapprox}) shows that the corresponding 
 amplitude of libration of the Tisserand frame $\mathbf K$ is  $1/6$ of the amplitude of libration of a rigid body.
  
\subsection{Energy dissipation rate}

The energy dissipation rate averaged over an orbital period $T=2\pi/n$
is given by (see \cite{RaR2017} appendix 2.1):
\[
    \frac{\Delta E}{T} =
    \frac{{\rm I}_{\circ}}{2}\frac{1}{T}\int_0^T
\!\ \mathrm{Tr}\!\ (\mathbf{F}\, \dot{\mathbf{B}}) dt.
  \]
Then equations  (\ref{oscparts}) and (\ref{Bij}) imply 
\begin{equation}\label{DtE}
  \frac{\Delta E}{T} =- n \,{\rm I}_\circ \, \Im\left[K_2(n)\right] \sum_{jk} |\hat F_{jk}|^2\,.
\end{equation}
Notice that the imaginary part of the Love number is negative, therefore the energy dissipated  $\Delta E$ is positive. 
From equation (\ref{Fij}) we obtain $\hat F_{11}=c_1e$, $\hat F_{22}=\hat F_{33}=-c_1e/2$,
$\hat F_{13}=\hat F_{31}=\hat F_{23}=\hat F_{32}=0$, and $\hat F_{12}=\hat F_{21}=\hat F_{12}=-c_1(ie+\hat\zeta)$.
Substituting these expressions into equation (\ref{DtE}) and replacing $c_1$ and $K_2(n)$ by
their expressions, we arrive at the following formula:
\begin{equation}\label{DtE2}
  \begin{split}
  \frac{\Delta E}{T}=\; & - \Im\left[k_2(n)\right]\frac{(nR)^5}{G}\left(\frac{m_2}{m_1+m_2}\right)^2
  \\ & \times
  \left[\frac{21}{2}e^2+12 e \Im(\hat\zeta) + 6 |\hat\zeta|^2\right]\;.
  \end{split}
\end{equation}

\subsection{Energy dissipation: comparison with other results in the literature}\label{compdiss}

Again the comparison of our results with others in the literature is not straightforward because we do not assume the body has an a priori
triaxial shape. So, in order to do the comparison we first suppose  that the body librates as if it were rigid with principal
moments of inertia $\overline A<\overline B<\overline C$. In this case the complex amplitude of libration $\hat \zeta$ is given by equation
(\ref{zetaapprox}) with $k_2(n)/k_2(0)=0$, namely $\hat\zeta=i\beta/2$, where $\beta$ is the classical  amplitude of libration
in equation (\ref{zetaefro}). Substituting this value of $\hat\zeta$ into equation (\ref{DtE2}) we obtain 
\begin{equation}
  \begin{split}
  \frac{\Delta E}{T}=\; & - \Im\left[k_2(n)\right]\frac{(nR)^5}{G}\left(\frac{m_2}{m_1+m_2}\right)^2
  \\ & \times
  \left[\frac{21}{2}e^2+6 e \beta + \frac{3}{2} \beta^2\right]\;,
  \end{split}
\end{equation}
which corresponds to the expression in \cite{EFR2018328} (equation (96)). 
This expression with  $\beta=0$ is that in  \cite{SEGATZ1988187} (equation (13)).



\section{Integration of the equations of motion}\label{sec_integration}

\subsection{Numerical settings}\label{settings}
In this section we specify all the constants and initial conditions necessary to integrate the equations of motion. The numerical integrations of the system of first order ODEs are carried out by a Runge-Kutta method of order $8$ with embedded error estimator of order $7$ due to Dormand \& Prince, with stepsize control \citep{HairerEDO1993}.


\subsubsection{\label{param}The choice of parameters}

The present-time physical and orbital parameters for Enceladus and Saturn are given in Table \ref{parenc}.
The  constant $\gamma$ is calculated using \citep{ragazzo2018theory}
\begin{equation}
\gamma=\frac{4}{5}\frac{Gm_{1}}{R_{\rm I}^{3}},\qquad {\rm I}_\circ=\frac{2}{5}m_{1}R_{\rm I}^{2},\nonumber 
\end{equation}
where $m_{1}$ is the mass and ${\rm I}_\circ$ is the mean moment of inertia of the satellite.

The physical dimensions of $\eta_0$ and $\alpha_0=1/J_0$ in equations (\ref{maineq1}) are  $1/\mathrm{s}$ and $1/\mathrm{s}^2$, respectively.
As discussed in  \cite{Correia2018},
if these constants are divided by 
\begin{equation}
\frac{152\pi}{15}\frac{R}{m_{1}}=7.43107\times 10^{-14}\,\mathrm{m}\cdot\mathrm{kg}^{-1},
\end{equation}
then the resulting parameters have the usual dimensions ($\eta_0[\mathrm{Pa}\cdot\mathrm{s}]$ and $\alpha_{0}[\mathrm{Pa}]$)
and represent the microscopic viscosity and rigidity of an equivalent homogeneous body (possibly with radius different from $R$)
with the same macroscopic
properties
(in this work at no time Enceladus is assumed to be homogeneous). Therefore these re-scaled constants
can be directly compared to those used
by \cite{efr2018}, namely $\alpha_{0}=10^{10}\,\mathrm{Pa}$ and $\eta_{0}=0.17\times 10^{14}\,\mathrm{Pa}\cdot\mathrm{s}$.
This gives the Maxwell time $\tau_{M}=\eta_{0}/\alpha_{0}=1700$\,s.
For the Andrade rheology we choose $\alpha=0.33$ and $\tau_{A}=\tau_{M}=1700\,\mathrm{s}$ \citep{SHK2013}. 
As discussed in Section \ref{aprox}, for this choice of parameters the number of Voigt elements in the extended Burgers model can be
taken as $n=6$. The rheological parameters associated with this extended Burgers model  are given in
Table \ref{par_and}. They were determined using  equation (\ref{jrtr}) (see \cite{Birk2010} for details).


\begin{table*}[h]
\begin{center}
\caption{The present-time physical and orbital parameters of Enceladus}

\begin{tabular}{c c c c}
\hline
Parameter \hspace{1.0cm} & Notation \hspace{1.0cm} & Units \hspace{1.0cm} & Values\\
\hline
Mass of the host (Saturn) \hspace{1.0cm} & $m_{2}$ \hspace{1.0cm} & $\mathrm{kg}$ \hspace{1.0cm} & $5.683\times 10^{26}$ \\
Mass  \hspace{1.0cm} & $m_{1}$ \hspace{1.0cm} & $\mathrm{kg}$ \hspace{1.0cm} & $1.08\times 10^{20}$ \\
Mean radius  \hspace{1.0cm} & $R$ \hspace{1.0cm} & $\mathrm{m}$ \hspace{1.0cm} & $252.1\times 10^{3}$\\
Eccentricity  \hspace{1.0cm} & $e$ \hspace{1.0cm} & \hspace{1.0cm} & $0.0045$ \\
Semi-major axis  \hspace{1.0cm} & $a$ \hspace{1.0cm} & $\mathrm{m}$ \hspace{1.0cm} & $2.38\times 10^{8}$ \\
Mean rotation rate  \hspace{1.0cm} & $n$ \hspace{1.0cm} & $\mathrm{rad}\cdot\mathrm{s}^{-1}$ \hspace{1.0cm} & $5.308\times 10^{-5}$\\
Orbital period \hspace{1.0cm} &  \hspace{1.0cm} & $\mathrm{s}$ \hspace{1.0cm} & $1.184\times 10^{5}$\\
Mean moment of inertia  \hspace{1.0cm} & ${\rm I}_\circ$ \hspace{1.0cm} & $\mathrm{kg}\cdot\mathrm{m}^{2}$ \hspace{1.0cm} & $2.2994\times 10^{30}$\\
Rigidity constant  \hspace{1.0cm} & $\gamma$ \hspace{1.0cm} & $\mathrm{s}^{-2}$ \hspace{1.0cm} & $4.69504\times 10^{-7}$\\
\hline
\end{tabular} \\[0.3em]
{\footnotesize  The mass of Saturn is taken from NASA Planetary Fact Sheets. The moment of inertia and the rigidity constant are calculated. All the other parameters are taken from Table 1 in \cite{efr2018}.}
\label{parenc}
\end{center}
\end{table*}


\begin{table}[h]
\begin{center}
\caption{Rheological parameters for the extended Burgers model.}

\begin{tabular}{c c c}
\hline
$r$ \hspace{1cm} & $1/\tau_{r} (\mathrm{yr}^{-1})$ \hspace{1cm} & $1/\eta_{r} (\mathrm{yr})$ \\
\hline

0 \hspace{1cm} & $18550.6$ \hspace{1cm} & $2.510112\times10^{-8}$ \\
1 \hspace{1cm} & $2378870$ \hspace{1cm} & $1.95734\times10^{-7}$ \\
2 \hspace{1cm} & $669017$ \hspace{1cm} & $8.89215\times10^{-8}$\\
3 \hspace{1cm} & $170566$ \hspace{1cm} & $3.88775\times10^{-8}$ \\
4 \hspace{1cm} & $37423.7$ \hspace{1cm} & $1.5879\times10^{-8}$\\
5 \hspace{1cm} & $6514.34$ \hspace{1cm} & $5.79894\times10^{-9}$\\
6 \hspace{1cm} & $782.589$ \hspace{1cm} & $1.7528\times10^{-9}$\\

\hline
\end{tabular} \\[0.3em]
{\footnotesize We use quantities in Table \ref{MKS} to change from MKS units to Year.}
\label{par_and}
\end{center}
\end{table}


\begin{table*}[h]
\begin{center}
\caption{Changing from MKS units to Astronomical unit, Mass of Sun and Year\label{MKS}}
\begin{tabular}{c c c c}
\hline
Parameter \hspace{1.4cm} & Notation \hspace{1.4cm} & Units \hspace{1.4cm} & Values\\
\hline
Astronomical unit\hspace{1.4cm} & $\mathrm{au}$ \hspace{1.4cm} & $m$ \hspace{1.4cm} & $149597871000$\\
Year\hspace{1.4cm} & $\mathrm{yr}$ \hspace{1.4cm} & $s$ \hspace{1.4cm} & $31536000$\\
Mass of Sun \hspace{1.4cm} & $M_{Sun}$ \hspace{1.4cm} & $kg$ \hspace{1.4cm} & $1.9885\times 10^{30}$ \\
\hline
\end{tabular}
\end{center}
\end{table*}


\subsubsection{Initial conditions for a Maxwell rheology}

The libration  of Enceladus, under Maxwell rheology and
moving on a fixed slightly eccentric orbit, is determined by equations  (\ref{moteqn1}) and (\ref{eqmax}).
In order to integrate these equations we use the following set of initial conditions:
\begin{equation}\label{incon}
\begin{split}
q'(0)&=\begin{pmatrix}
0.001598\\
0\\
0
\end{pmatrix}\,\mathrm{au},
v'(0)=\begin{pmatrix}
0\\
2.649\\
0
\end{pmatrix}\, \mathrm{au} \cdot \mathrm{yr}^{-1},\\
\mathbf{\omega}(0)&=\begin{pmatrix}
0 & -1672.44 & 0\\
1672.44 & 0 & 0\\
0 & 0 & 0
\end{pmatrix}\, \mathrm{yr}^{-1},\\
\mathbf{b}(0)&=\begin{pmatrix}
0.01398 & 0 & 0\\
0 & -0.00399 & 0\\
0 & 0 & -0.00998
\end{pmatrix}.
\end{split}
\end{equation}


\subsubsection{Initial condition for an Andrade rheology}

The libration  of Enceladus under Andrade  rheology  is determined by equations  (\ref{moteqn1}) and (\ref{eq11}). The parameters of
the rheology are those in Section \ref{param} and the initial conditions are those in equation (\ref{incon}).
The Andrade model also requires the initialisation of the internal degrees of freedom of the rheology, namely
$\mathbf{b}_1,\mathbf{b}_2,\ldots$. At time $t=0$ they are all  set to zero.


\subsection{Results}

\subsubsection{Forced libration}
\label{sec:lib}

\begin{figure}[ptb]
\centering
\includegraphics[scale=0.6]{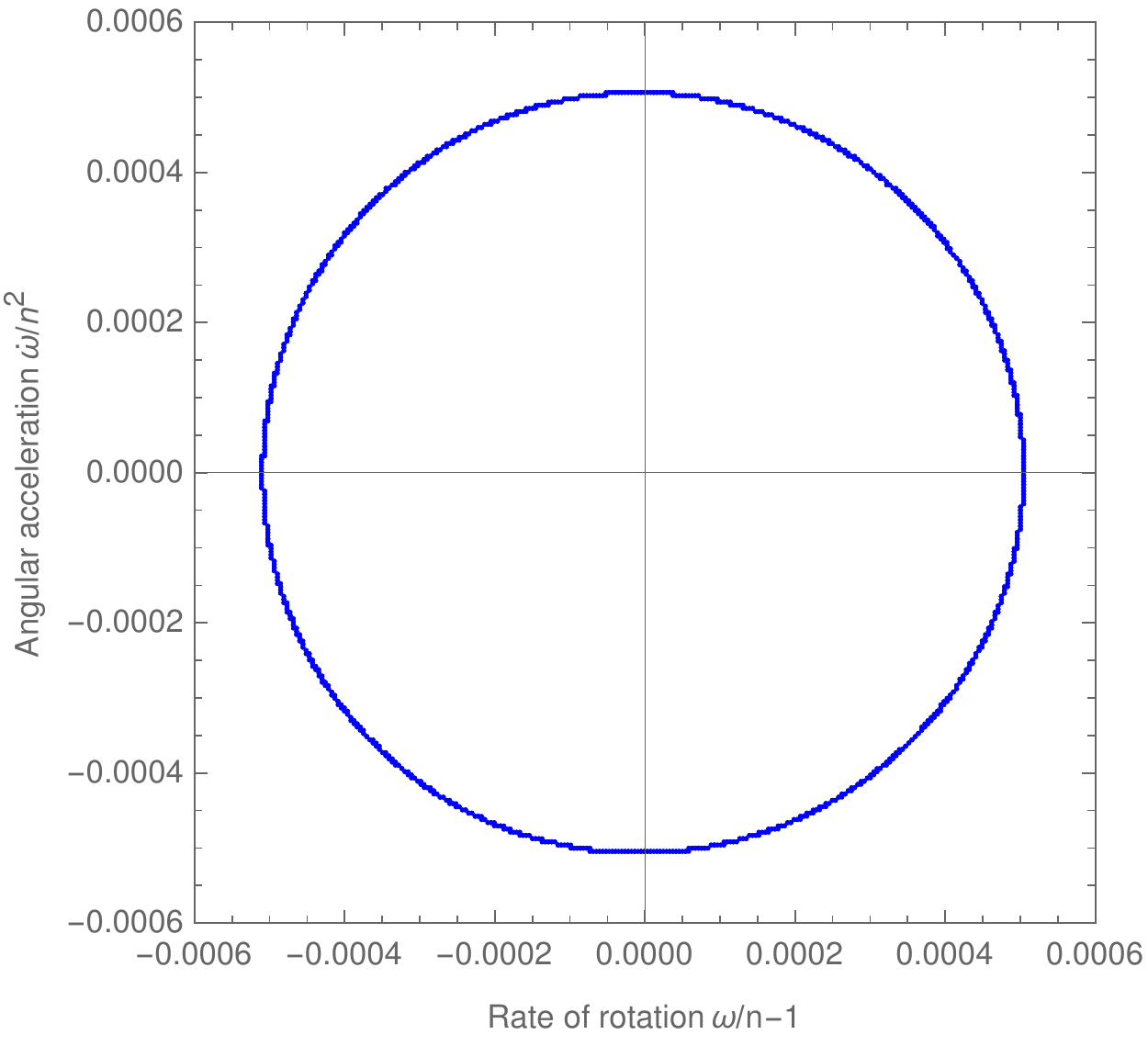}
\caption{Libration in longitude of Enceladus obtained by numerical integration of the equations of
motion (\ref{moteqn1}) and (\ref{eq11}) associated with the extended Burgers model with 6 Voigt elements. The evolution is plotted in terms of angular acceleration ($\dot{\omega}/n^{2}$)
versus normalised spin rate $(\omega-n)/n$ after the decay of the transient state. The same figure
obtained in the Maxwell case from the integration of Eqs.~ (\ref{moteqn1}) and (\ref{eqmax})
is not represented because the difference between the two models is smaller than the thickness of
the curve.\label{fig:acc_and}}
\end{figure}

\begin{figure}[ptb]
\centering
{\small\color{black!70} \qquad\qquad Maxwell rheology} \\[0.5em]
\includegraphics[scale=0.6]{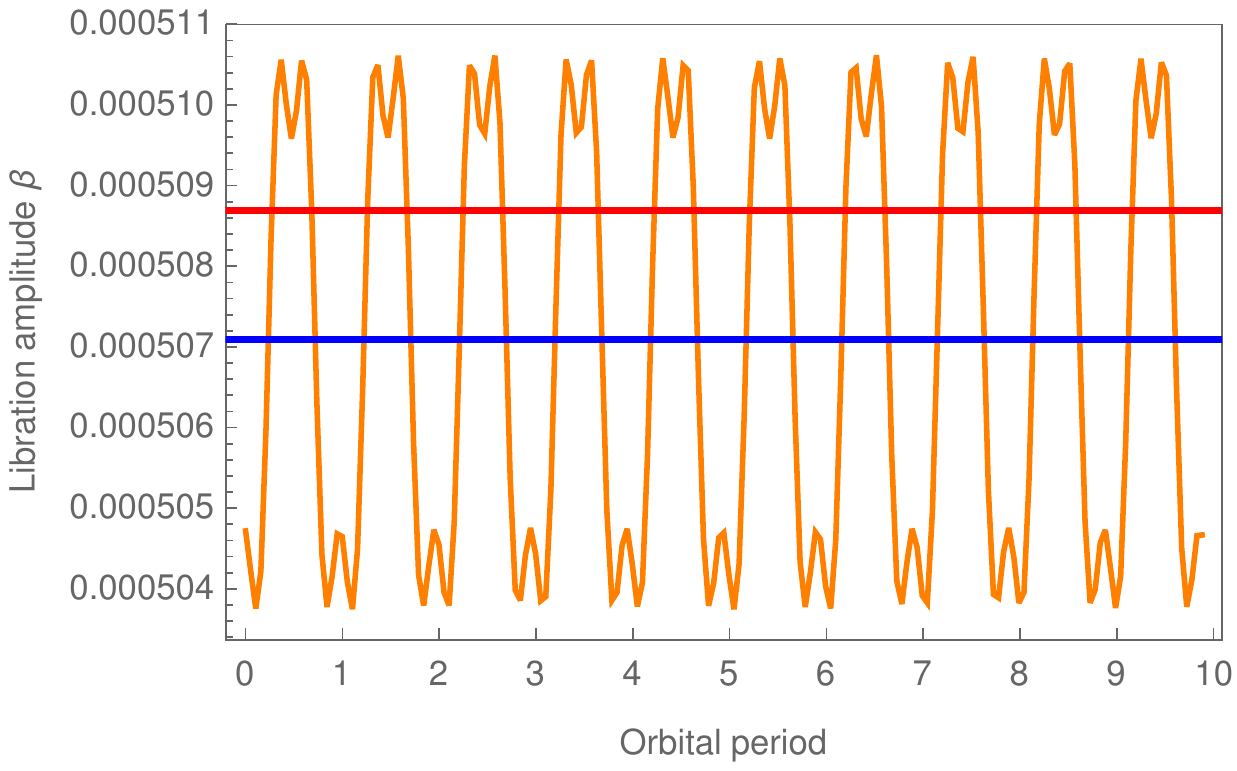} \\[0.5em]
{\small\color{black!70} \qquad\qquad Extended Burgers model with 6 Voigt elements} \\[0.5em]
\includegraphics[scale=0.6]{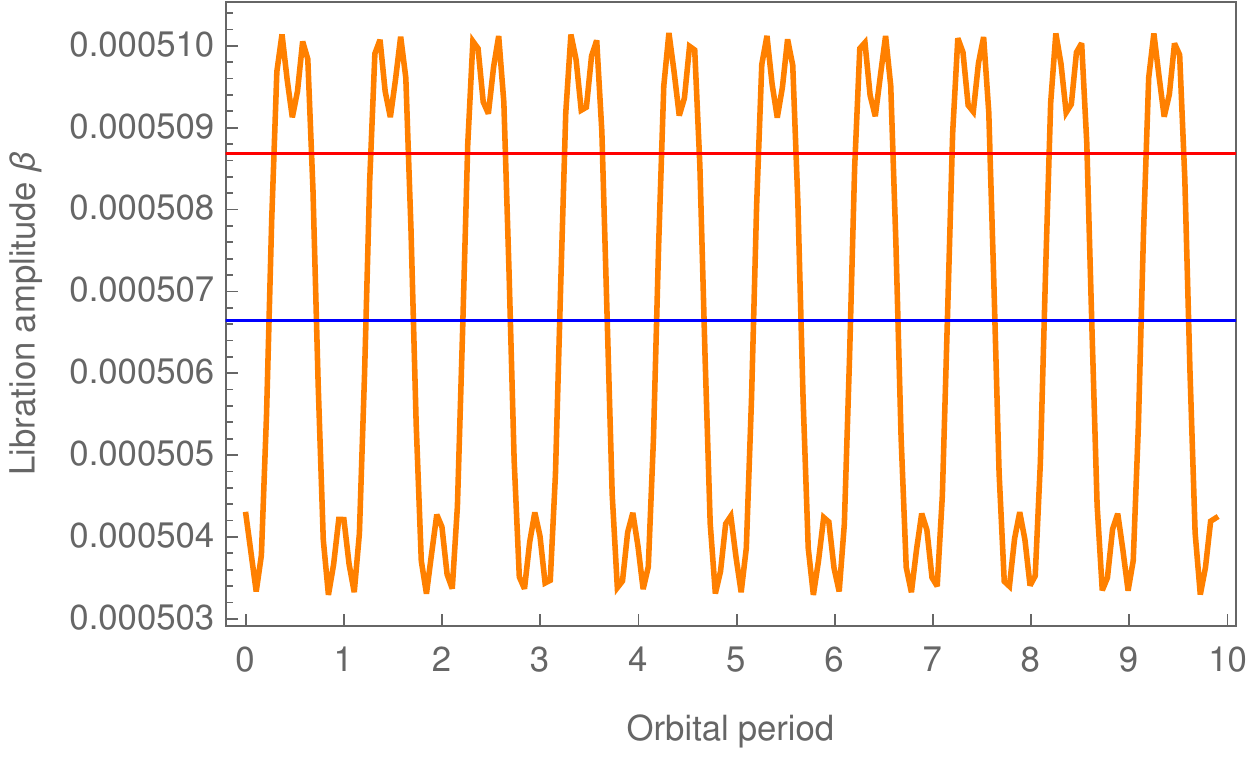}
\caption{Amplitude of libration in longitude after the decay of the transient state. The orange curve is obtained by numerical integration of the equations of motion (\ref{moteqn1}) and (\ref{eqmax}) in the Maxwell case (top) or (\ref{moteqn1}) and (\ref{eq11}) in the Andrade case (bottom). The blue horizontal lines show the mean values of the integrations. The red horizontal lines represent the analytical solutions (\ref{betaapprox}).\label{fig:lib_amp}}
\end{figure}

The rotation angle
$\theta$ does not appear  explicitly in the equations of motion in the inertial reference frame,
(\ref{moteqn1}), (\ref{eqmax}),  and (\ref{eq11}), though $\dot \theta=\Omega_3$ does.  
The main component of the forced
libration $\zeta = \theta - M$ is oscillating at the orbital frequency $n$ (see Section~\ref{sec_libration}) and 
once the free libration is totally damped the amplitude of forced libration is given by 
\begin{equation}\label{eq.numbeta}
  \beta=
  \sqrt{\left(\frac{\Omega_3}{n}-1\right)^2 + \left(\frac{\dot\Omega_3}{n^2}\right)^2}\;.
\end{equation}
The two quantities in the square root, obtained by numerical integration of the equations of motion, are plotted against each other
in Figure~\ref{fig:acc_and}. The circular trajectory described in this set of coordinates justifies
the expression (\ref{eq.numbeta}). The associated radius $\beta$ is plotted in Figure~\ref{fig:lib_amp}
together with the analytical approximations derived in Section~\ref{sec_libration}. As expected from
Section~\ref{number}, at the orbital frequency of Enceladus the amplitudes of libration are very close for
both Maxwell and Andrade rheologies. We obtained a value of 0.000508\,rad which is about one fourth of the
observed amplitude. This apparent contradiction is discussed in Section~\ref{difference}. The results are in good agreement with the analytical
approximation (\ref{betaapprox}). Nevertheless, the numerical outputs show a small modulation of this amplitude not
captured by the analytical approximation and a slight offset of $0.4\%$ in both the Maxwell and the Andrade cases. But for Enceladus these differences have negligible consequences on the
averaged dissipation.


\subsubsection{Dissipation rate}

\begin{figure}[ptb]
\centering
{\small\color{black!70} \qquad\qquad Maxwell rheology} \\[0.5em]
\includegraphics[scale=0.6]{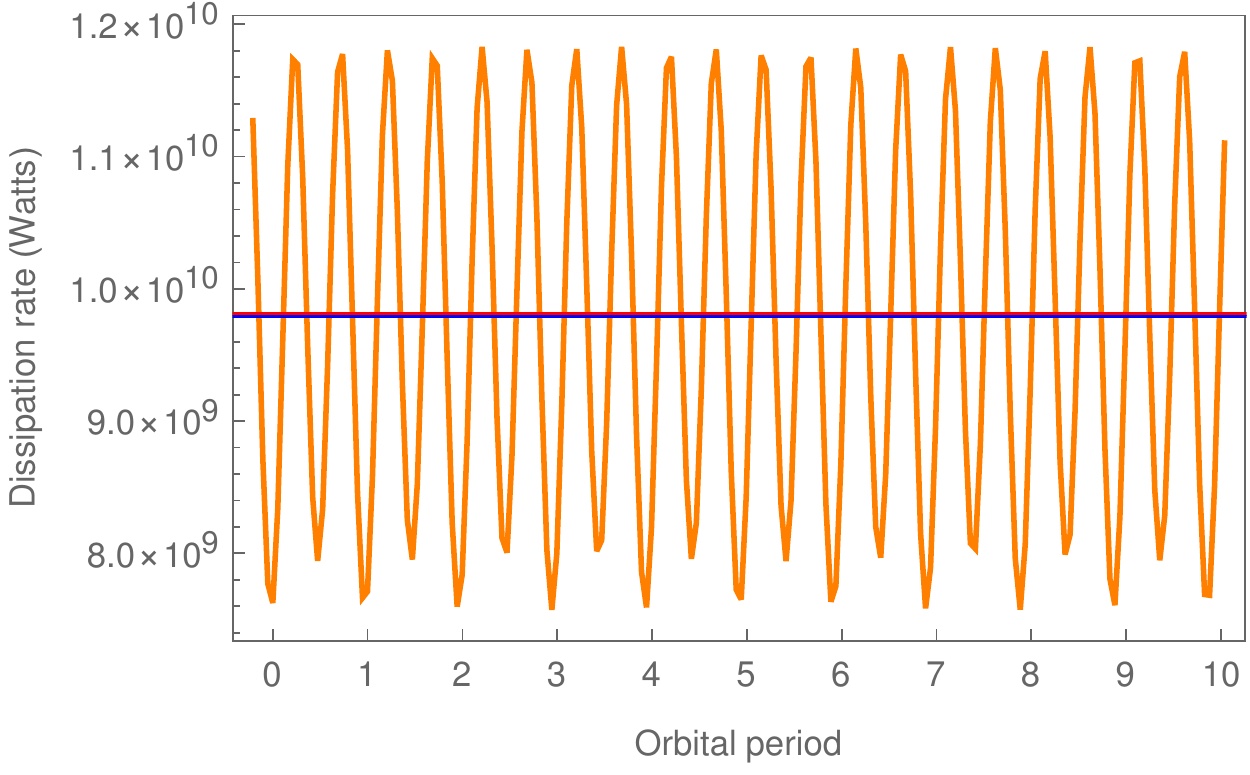} \\[0.5em]
{\small\color{black!70} \qquad\qquad Extended Burgers model with 6 Voigt elements} \\[0.5em]
\includegraphics[scale=0.6]{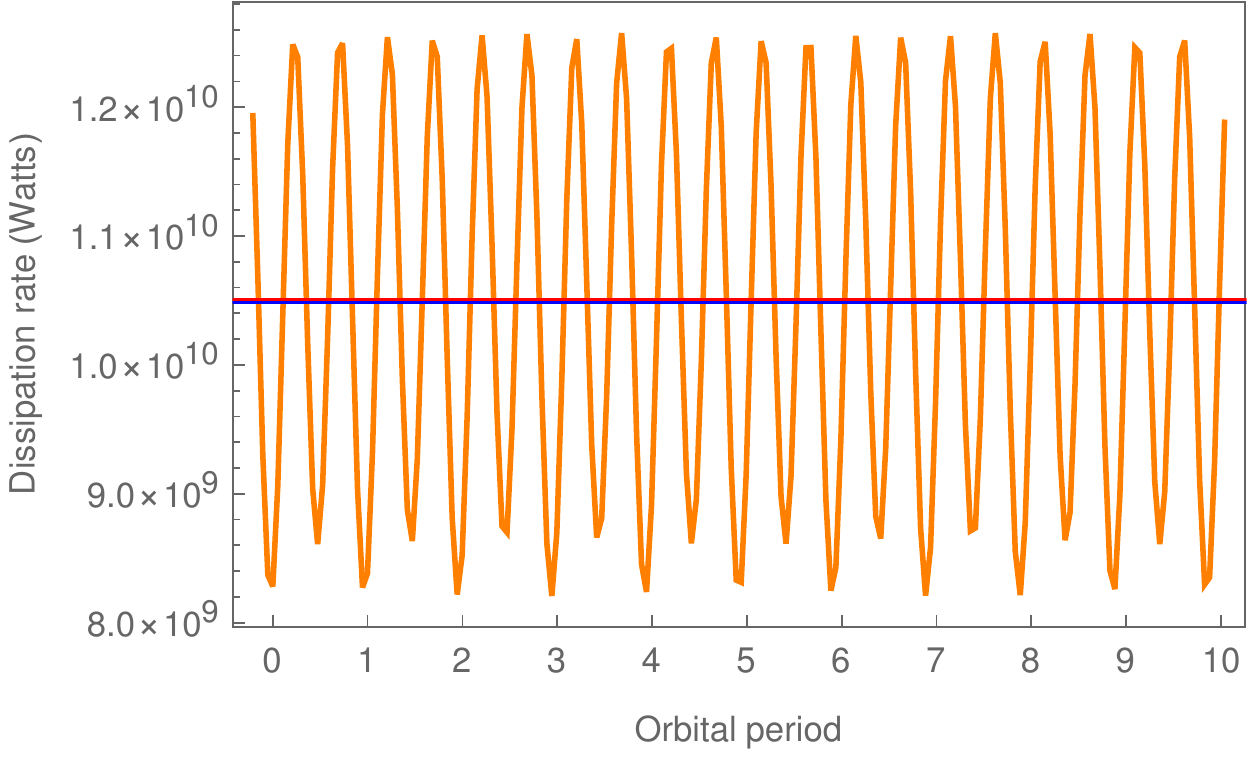}
\caption{Dissipation rate (in Watts) as a function of time in orbital period once the steady state
is reached. The orange curve is obtained by numerical integration of the equations of motion (\ref{moteqn1}) and (\ref{eqmax}) in the Maxwell case (top) or (\ref{moteqn1}) and (\ref{eq11}) in the Andrade case (bottom).
The blue horizontal lines show the mean values of integration. The red horizontal lines represent the analytical expression (\ref{DtE2}). \label{fig:dis_fix_max}}
\end{figure}

\begin{figure}[ptb]
\centering
\includegraphics[scale=0.35]{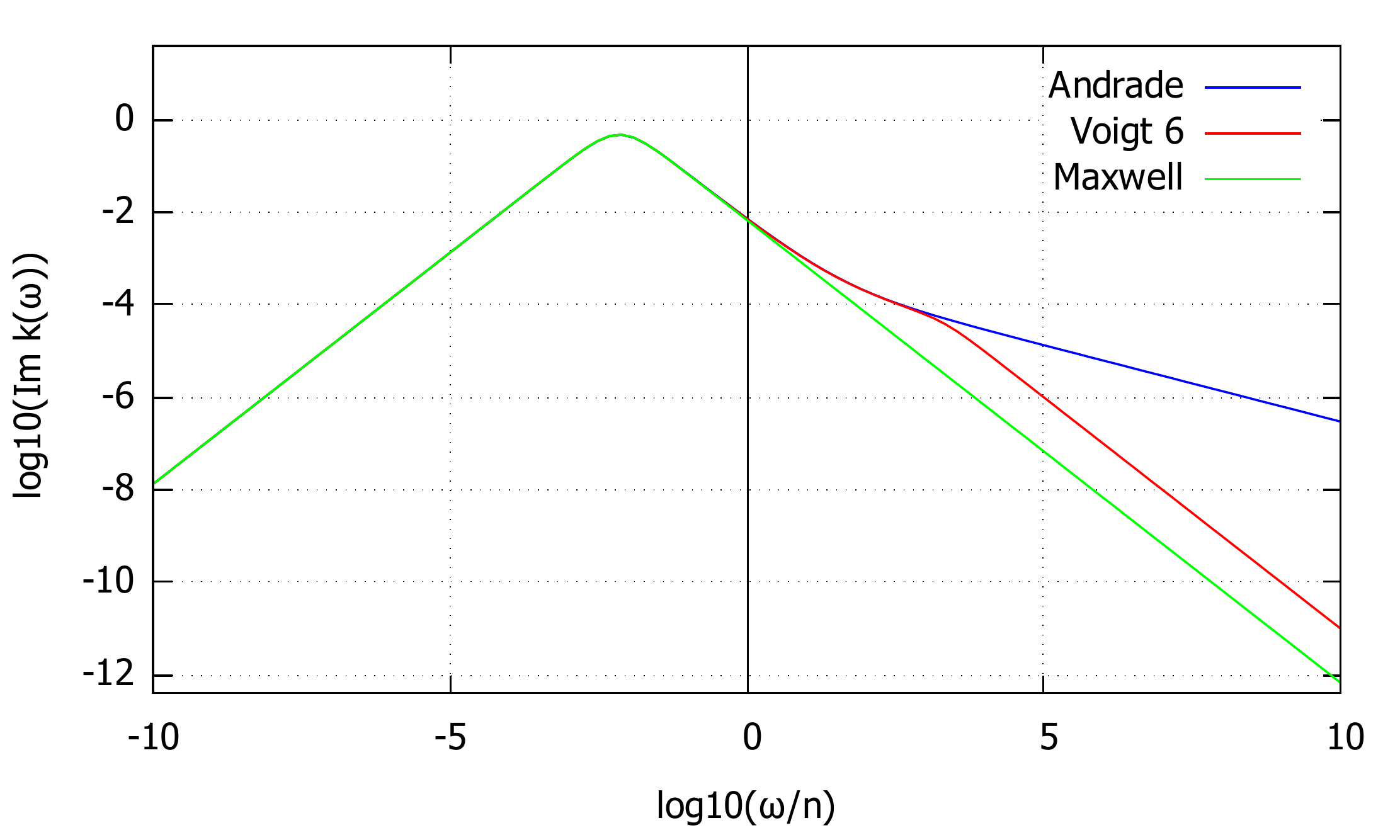}
\caption{Imaginary part of the Love number as a function of $\omega$. The black vertical line marks the value of mean rotation rate of Enceladus.\label{fig:k_2(w)}}
\end{figure}

With the numerical integrations, we have access to the instantaneous dissipation rate within the
system given by the sum of the dissipation rates in all dash-pots of the oscillator
representing the rheology. Results are plotted in Figure~\ref{fig:dis_fix_max} over 10 orbital
periods after the damping of the free libration. They display small oscillations with amplitude of 2\,GW 
at twice the orbital frequency around a mean value equal to 9.80\,GW and 10.50\,GW in the
Maxwell and in the extended Burgers model, respectively. The closeness of these two values is due to
the fact that the imaginary part of the Love number is about the same in both rheologies at the
orbital frequency (see Figure~\ref{fig:k_2(w)}).

For comparison, the analytical dissipation rates given by equation~(\ref{DtE2}) are respectively
9.81\,GW and 10.51\,GW. These values, represented as horizontal blue lines in
Figure~\ref{fig:dis_fix_max}, are in very good agreement with the averaged numerical values.

Given the low amplitude of the forced libration in our model compared to the observed one (see
previous section), their contribution to the total dissipation rate is only of the order of 7\%
instead of the 30\% previously reported based on the observed libration \citep{efr2018, Folonier2018}.


\section{Observed vs Model libration: should they be the same?}

\label{difference}

 The observed magnitude  of libration  of Enceladus is $0.0021$ rad
 (see \cite{THOMAS201637}). The value computed in this paper, with rheological parameters from the literature,
 is $0.000508$ rad.  \footnote{\label{footporco}In \cite{THOMAS201637} (Table 1)  the amplitude of libration $\beta$ of Enceladus  was
   computed for four different interior models:
  - one  homogeneous  ellipsoidal rigid body in hydrostatic equilibrium, - two  different two-layers rigid bodies in
  hydrostatic equilibrium, and - an external ice shell over a  subsurface ocean over a solid core. The value of $\beta$ found for
 the first three models was approximately $0.033^\circ$. 
  The value of $\beta$ found in our paper, for a deformable Enceladus with two
  different rheologies and no a priori assumed deformation,  is approximately $0.03^\circ$. The agreement between our result and that in
  \cite{THOMAS201637}   is very good and the difference may be due to the
  fact that they used the rigid body formula as that in equation (\ref{zetaefro}) while we used (\ref{zetae}) that takes into account
  the  rheology of the satellite. 
  The disagreement of all these values of $\beta$   with the
  observed amplitude of libration of Enceladus' surface, $0.120^\circ$,   was the main evidence found in  \cite{THOMAS201637} for
  the existence of a subsurface ocean. Within the context of our model we present another,  but equivalent,
  explanation for this difference, see text.}Should these figures  be the same? The answer is no, because they  measure  different
 physical entities.
In this paper the Enceladus' frame $\mathbf{K}$  
 is a Tisserand frame, namely, it  
 is an orthogonal moving frame with the origin at the centre
of mass of Enceladus and with an angular velocity with respect to the inertial frame such that
the angular momentum of the body with respect to $\mathbf{K}$ is null \citep{MunkMac}. 
While in this paper we computed the libration of $\mathbf K$, the
observed magnitude  of libration  of Enceladus is the libration of its external shell that
moves over a global subsurface sea  \citep{THOMAS201637}.

In order to relate the libration of   $\mathbf K$ to that of the shell we will assume, as done in \cite{THOMAS201637}, an interior
model for  Enceladus made of three layers.  The external layer is a spherical
shell 23\,$\mathrm{km}$  thick with density 850\,$\mathrm{kg}/\mathrm{m}^3$. The second layer 
is a spherical ocean 28.1\,$\mathrm{km}$ thick
with density 1000\,$\mathrm{kg}/\mathrm{m}^3$. The third layer is a spherical core of density  2300\,$\mathrm{kg}/\mathrm{m}^3$. The
mean radius of Enceladus is 252.1\,$\mathrm{km}$. Let $\dot \theta_1$, $\dot \theta_2$, and $\dot \theta_3$ denote the
angular velocity (in the Tisserand sense) of each layer with respect to the inertial frame.
If ${\rm I}_1$, ${\rm I}_2$,  and ${\rm I}_3$ denote the moment of inertia of each layer
then ${\rm I}_1+{\rm I}_2+{\rm I}_3={\rm I}_\circ$  and the definition of Tisserand's frame implies
\begin{equation}
  {\rm I}_1(\dot \theta_1-\dot \theta)+ {\rm I}_2(\dot \theta_2-\dot \theta)+{\rm I}_3(\dot \theta_3-
  \dot \theta)=0\label{libshell} \, .
\end{equation}
Supposing that the liquid ocean does not librate, $\dot \theta_2=n$,  and that the angle of libration of the
shell and core are denoted as $\zeta_1$ and $\zeta_3$,  with $\dot \theta_1=n+\dot\zeta_1$ and
$\dot \theta_3=n+\dot\zeta_3$, then equation (\ref{libshell}) implies
\begin{equation}
  {\rm I}_1\dot \zeta_1+ {\rm I}_3\dot \zeta_3={\rm I}_\circ\dot \zeta \label{libshell2} \,.
\end{equation}
Assuming that all librations are in phase then this equation can be integrated to give
\begin{equation}
  {\rm I}_1 \zeta_1+ {\rm I}_3\zeta_3={\rm I}_\circ \zeta \label{libshell3} \,.
\end{equation}
The substitution of the  values of ${\rm I}_1$, ${\rm I}_3$, ${\rm I}_\circ$,
$\zeta_1=0.0021$, and $\zeta=0.000508$ into equation (\ref{libshell3}) gives an estimate 
$\zeta_3=0.006\, \zeta_1$. So, assuming the interior model of Enceladus given in \cite{THOMAS201637}
then the results in this paper imply that the amplitude of the libration of the  core is approximately
$0.6\%$ of that of the  shell.


\section{Conclusions}\label{conclusions}
In this paper, we present a method to simulate the evolution in the time domain of an extended body
subject to tides with arbitrary rheology. In particular, Andrade's creep model is known to be well
suited for modelling tidal deformation of icy satellites. But it involves a fractional power law in
the frequency domain preventing a straightforward transposition into a set of ordinary differential
equations in the time domain. To circumvent this issue, Andrade's compliance function is
approximated by that of an extended Burgers model containing an arbitrary number of Voigt elements.

The procedure is applied to Saturn's satellite Enceladus for which the forced libration in longitude
has been shown to be responsible for a significant amount of energy dissipation. For this particular
body, we consider both the Maxwell and the Andrade visco-elastic models. Tidal effects on the orbit
are switched off in order to maintain a constant eccentricity which, in the special case of
Enceladus, is driven by the 2:1 mean motion resonance with Dione.

In this particular setting, and for arbitrary rheology represented by a Love number $k_2$, we show
that the amplitude of libration $\beta$ is given by
\[
\beta \approx 6 e \frac{\overline B - \overline A}{\overline C} \frac{m_2}{m_1+m_2}
\left|1-\frac{5}{6}\frac{k_2(n)}{k_2(0)}\right|
\,.
\]
This equation predicts values reduced by a factor $|1-5k_2(n)/6k_2(0)|$ with respect to the classical
formula obtained in the case of a strictly rigid body with the same permanent triaxiality.
We also recover the expression of the average rate of tidal dissipation of a synchronous body valid for any rheology,
namely \citep{EFR2018328}
\[
\begin{split}
\frac{\Delta E}{T} = & -\Im[k_2(n)] \frac{(nR)^5}{G} \left(\frac{m_2}{m_1+m_2}\right)^2
\\ & \times
\left[\frac{21}{2}e^2+6e\beta+\frac{3}{2}\beta^2\right]\, .
\end{split}
\]
Thanks to these two formulae, we have been able to validate the numerical procedure. The analytical
estimates of the libration amplitude (0.0005\,rad) and of the dissipation rate (10\,GW) differ by
less than 0.4\% from the averaged values obtained by numerical integration. In our model, the
libration in longitude is that of the Tisserand frame. The comparison with the observed libration
of the shell allowed us to estimate that the core librates with an amplitude 0.6\% times smaller
than the shell.

In addition to the averaged libration amplitude and dissipation rate, the numerical integration also
displays oscillations not captured by the first order analytical calculation.  Moreover, the
procedure can be applied in more complex situations where a Fourier series is not feasible, e.g., 
in systems with large eccentricities and/or with non quasiperiodic perturbations as in a chaotic
system with close encounters.


\section*{\label{ack}Acknowledgments}

We would like to thank Michael Efroimsky and another anonymous referee for careful reading and many suggestions improving the manuscript. CR is partially supported by FAPESP grant 2016/25053-8. YG is partially supported by FAPESP grants 2015/26253-8 and 2018/02905-4. AC is supported from CFisUC strategic project (UID/FIS/04564/2019), ENGAGE SKA (POCI-01-0145-FEDER-022217), and PHOBOS (POCI-01-0145-FEDER-029932), funded by COMPETE 2020 and FCT, Portugal. 


\end{document}